\documentclass[12pt]{article}
\usepackage{graphicx,psfrag,epsf}
\usepackage{url} 
\usepackage{geometry, booktabs, subcaption}
\usepackage{titlesec}
\usepackage[T1]{fontenc}
\usepackage[utf8]{inputenc}
\usepackage{authblk}
\usepackage{amssymb,amsmath}
\usepackage{graphics}
\usepackage[english]{babel}
\usepackage{amsmath,amssymb,enumerate,rotate,multirow}
\usepackage{hyperref}
\usepackage{bm,latexsym,mathrsfs}
\usepackage[usenames]{color}
\usepackage{graphicx}
\usepackage{subcaption}
\usepackage[authoryear]{natbib}
\usepackage{booktabs}
\usepackage{colortbl}
\usepackage{xcolor}
\usepackage{xfrac}
\usepackage{todonotes}

\usepackage{xspace}
\usepackage{amsthm}
\usepackage{bbm}
\usepackage{verbatim}
\usepackage{multirow}

\usepackage{sidecap} 
\sidecaptionvpos{figure}{c}
\usepackage{pdflscape}

\usepackage{placeins}
\usepackage[figuresleft]{rotating}

\newcommand{\calB}{\mathcal{B}}
\newcommand{\calN}{\mathcal{N}}

\newcommand{\invWish}{\mathcal{IW}}
\newcommand{\Law}{\mathcal L}

\newcommand{\vvec}{\text{vec}}
\newcommand{\be}{\begin{equation}}
\newcommand{\ee}{\end{equation}}
\newcommand{\iid}{\stackrel{\mathrm{iid}}{\sim}}
\newcommand{\ind}{\stackrel{\mathrm{ind}}{\sim}}

\newcommand{\Det}[1]{\vert #1 \vert}

\DeclareMathOperator{\tr}{tr}

\newcommand{\virgolette}[1]{``#1"}

\usepackage[authoryear]{natbib}

\allowdisplaybreaks

\usepackage{xr} 

\makeatletter
\newcommand*{\addFileDependency}[1]{
  \typeout{(#1)}
  \@addtofilelist{#1}
  \IfFileExists{#1}{}{\typeout{No file #1.}}
}
\makeatother

\begin{document}

\title{Bayesian Nonparametric Vector Autoregressive\\ Models
 via a Logit Stick-breaking Prior: \\an Application to Child Obesity}

\author{ Mario Beraha$^{1, 2}$, Alessandra Guglielmi$^2$, Fernando A. Quintana$^3$,  \newline Maria de Iorio$^4$, Johan Gunnar Eriksson$^4$ and Fabian Yap$^5$\\
$^1$ Universit\`{a} di Bologna, $^2$ Politecnico di Milano,\\
$^3$ PUC Santiago de Chile, $^4$  NUS Singapore\\ and $^5$ KK Women's and Children's Hospital}
\date{\today}
\maketitle

\begin{abstract}
Overweight and obesity in adults are known to be associated
with risks of metabolic and cardiovascular diseases.
Because obesity is an epidemic, increasingly affecting children, it is important to understand if this condition persists from early life to childhood and if different patterns of obesity growth can be detected. Our motivation starts from a study of obesity over time in children from South Eastern Asia. Our main focus is on clustering
obesity patterns after adjusting for the effect of baseline
information. Specifically, we consider a joint model for height and weight patterns taken every 6 months from birth. We propose a novel model that facilitates clustering by combining a vector autoregressive sampling model with a dependent logit stick-breaking prior. Simulation studies show the superiority of the model to capture patterns, compared to other alternatives. We apply the model to the motivating dataset, and discuss the main features of the detected clusters. We also compare alternative models with ours in terms of predictive performances. 
\end{abstract}

\section{Introduction}
\label{sec:intro}

Overweight and obesity are defined as abnormal or excessive fat accumulation that may impair health \citep{WHO}. It is well-known that overweight and obesity in adults are associated with  risks of metabolic and cardiovascular
diseases; see, for instance, \cite{doi:10.1161/ATVBAHA.107.159228},  \cite{doi:10.1161/CIRCULATIONAHA.106.675355} and \cite{pi2009medical}. Obese individuals are also associated with a more severe course of illness with COVID-19 \citep{gao2020obesity}.

Obesity is an epidemic, increasingly affecting children. In 2018, 18\% of children in the United States were obese and approximately 6\% were severely obese \citep{hales2018trends}. Prevalence of obesity
in children has increased from 4\% in 1975
to over 18\% in 2016 among children and adolescents aged 5-19 years [WHO,
Accessed: 01-06-2021]; see also \cite{cremaschi2021integrating}. Overweight or obesity in childhood is critical as it often
persists into adulthood due to both physiological and behavioural factors, e.g. (i) adults diet based on energy-dense foods that are high in fat and sugars and (ii) adult physical inactivity due to the  sedentary nature of many forms of work, changing modes of transportation, and increasing urbanization.
Also for childhood obesity, dietary composition and sedentary lifestyle have often been cited as main contributors. Evidence also exists for a significant role of parents’ socioeconomic status and maternal prenatal health factors; see \cite{cremaschi2021integrating}.

Research on the origins of health and disease suggests that susceptibility to metabolic disease may originate early in life. Different conditions in maternal uteruses seem to influence metabolic health by
altering glucose metabolism and body composition. See \cite{symondsetal:13} and \cite{10.1371/journal.pone.0041759}.
Moreover, increased adiposity have been observed in school-age children and infants~\citep{10.1093/ije/dyq180, Whincup:05, 10.1210/jc.2002-020434, yajnik:03}.

It is therefore important to understand whether obesity persists from early life to childhood and if different types of obesity growth can be detected. For instance, \cite{zhang2019rate} show that rates of change in Body Mass Index (BMI)  at different childhood ages are differentially associated with adult obesity. 
Our motivating application is the study of obesity over time in a dataset of children in South Eastern Asia \citep[see][]{GUSTO}, taken every 6 months from birth. In particular, we consider both their height and weight.  It is known that obesity might increase the risk of metabolic diseases, and that this risk is higher in Asian populations than in White Caucasian population \citep{misra2011obesity}. 
 The aim of this manuscript is to cluster children according to obesity growth patterns. Information is available on children as well as mothers. In particular, we focus on clustering  the children after adjusting for covariates (of both fixed and time-varying types). We assume  a bivariate vector autoregressive (VAR) model for
joint responses (height and weight). VAR models may provide a flexible and powerful representation of longitudinal data;
see, for instance, \cite{canova2004forecasting} and \cite{daniels2002bayesian}. 
 
We specify a time-dependent Bayesian nonparametric prior on the VAR coefficients to allow for data-driven clustering of the children.  More in details, we assume the children-specific VAR coefficients to be  independently distributed according to a truncated stick-breaking prior with weights that depend on  baseline covariates. This construction induces a prior on the partition of the children in the sample. Moreover, it allows for potentially empty
clusters, in which case the \textit{number of clusters} is interpreted as the
number of non-empty components in the stick-breaking representation, i.e. components to which at least one observation is assigned.

The dependent stick-breaking prior adopted here can be seen as a finite-dimensional version of the 
logistic stick-breaking process described in \cite{ren2011logistic}. Covariate dependent random probability constructions include the probit stick-breaking process \citep{chung2009nonparametric,rodriguez_dunson_2011}. These 
Bayesian nonparametric random probability measures stem out from the seminal work by \cite{maceachern2000dependent} on dependent Dirichlet processes. See a review of this and related models in \cite{QuMuJaMa2022}. 
Covariate dependent priors for  random partition  were first proposed in \cite{muller2011product} and \cite{park2010bayesian}.

Bayesian nonparametric methods have  been successfully applied to VAR models in recent years. See \cite{kalli2018bayesian} for such a model applied to 
single subject data,
\cite{billio2016bayesian} and  \cite{kundu2021non} for multiple subject data. 
In \cite{billio2016bayesian} the authors
propose a Dirichlet process
mixture of normal-Gamma priors on the VAR autocovariance elements, as a Bayesian-Lasso prior.  \cite{kundu2021non} focus on matrix-variate data, providing a  class of non-parametric Bayesian VAR
models, based on heterogeneous multi-subject data, that enables separate clustering at multiple
scales, and result in partially overlapping clusters. 
 
Our contribution  includes the design of an efficient Gibbs sampling algorithm to perform posterior inference, 
 that exploits recent results on logit stick-breaking priors by \cite{rigon_durante_logit}. Note that the this random probability measure is represented by a finite random probability with $H$ support points, but, unlike the sparse mixture in \cite{fruhwirth2019here}, (i) the weights depend on covariates and (ii) come from a stick-breaking construction, thus implying stochastic dominance of the sequence itself (for a fixed value of the covariate).  

Finally,  \cite{cremaschi2021integrating} consider a more complex model in a similar framework, i.e. they  provide a joint model for multiple
growth markers and metabolic associations, which allows for data-driven clustering
of the children and highlights metabolic pathways involved in child obesity.  Unlike our approach,
they  assume  a Bayesian joint  non-parametric
random effect distribution on the parameters characterizing the longitudinal
trajectories of obesity and the graph capturing the association between
metabolites.

The remainder of this paper is structured as follows.   Section~\ref{sec:data}
describes the motivating application and introduces a preliminary exploratory analysis.
In Section~\ref{sec:model} we present the finite mixture of VAR models
and discuss its main features. Section~\ref{sec:simulation} summarizes the results of three simulation studies
carried out to test and compare posterior inference   under possible alternative model formulations.  
Section~\ref{sec:datanalysis} presents the results from the main application. Section~\ref{sec:discussion} concludes the paper with a discussion. 
The appendix provides details on the Gibbs sampler algorithm for
posterior simulation, and presents further simulations.

\section{Child growth dataset}
\label{sec:data}

We focus on the analysis of obesity in children from Singapore, particularly on its evolution over time.
As mentioned in the Introduction, it is relevant to understand whether obesity persists from early life to childhood.
  Such information is of particular relevance when designing intervention policy.
Section~\ref{sec:data_description} introduces the data and explains the main research questions,
while Section~\ref{sec:EDA} contains a short summary of
the exploratory analysis carried out to highlight the main data characteristics.  

\subsection{Description of the dataset}
\label{sec:data_description}

We consider data from \textit{the Growing Up in Singapore Towards healthy Outcomes} (GUSTO) study, which  comprises one of the most carefully phenotyped parent-offspring cohorts with a particular focus on epigenetic observations; see \cite{GUSTO} for description of the recruited women and objectives of the cohort study. 
The  data consist of measurements of child height (or length, depending on the child's age) in centimeters  and weight in kilograms from periodic visits of $1139$ children
from  birth  to the age of seven. We consider only visits occurred every 6 months, though during the first year of life, infants were visited every 3 months. 
More specifically, the response vector $\bm y_{it} \in \mathbb{R}^2$ is given by the  measurements of (\textit{length}, \textit{weight}) up to the 12th month of age ($t=3$) and 
(\textit{height}, \textit{weight}) from the 18th month  onwards ($t=4, \ldots, 14$).
Besides sex of the child, information is available on the mother. However, the original sample includes missing observations. More in details, 77 subjects are discarded from the analysis, because only  information on the first visit (i.e. right after birth) is available. Moreover, we discard 
children with less than two consecutive visits, and with missing baseline covariates. This leads to a final  sample size  of  $N=766$.  Note that we keep children with missing responses, since in Bayesian framework it is straightforward to imputing these as part of the MCMC. To this end, we simulate  the missing responses from their full conditional distribution at every iteration of the algorithm. See the MCMC algorithm in  Appendix~\ref{sec:gibbs}. 
 
The available  baseline covariates in the dataset 
 are:
\begin{itemize}
    \item \textit{age}, mother's age:   it ranges from 18 to 46 years.
    \item \textit{parity}: number of previous
pregnancies carried to a viable gestation  by the mother, ranging from 0 to 5.  If  parity equals to 0, the child is the first born.
    \item \textit{OGTT fasting Pw26}: oral glucose tolerance test  (OGTT) at  24th-26th week of pregnancy; it varies from 2.9 to 8.7 mg/dL.  Mothers are tested  after fasting for at least eight hours.    
    \item \textit{OGTT 2hour Pw26}: oral glucose tolerance test at 24th-26th week of pregnancy; it  ranges from 2.9 to 15.1 mg/dL.  
Mothers are tested two hours after having assumed  a glucose solution containing a dose 
of sugar. 
    \item \textit{ppBMI}: pre-pregnancy body mass index of the mother; values in the sample range from 14.6 to 41.3 Kg/m$^2$.
    \item \textit{GA}: gestational age in weeks, i.e. the length of the pregnancy (from 28 to 41.4 in the dataset).
    \item \textit{sex}: sex of the child.
    \item Mother's \textit{ethnicity}:  Chinese, Malay or Indian with proportions reflecting those characterising the  Singaporean population.  
    \item Mother's highest \textit{education}:  it is a categorical variable with three ordered  levels. Level 1 corresponds to no education or primary school, level 2 corresponds either to primary school, GCE (Singapore-Cambridge general certificate of education (O-level)) or ITE NTC (institute of technical education, national technical certificate) and level 3 corresponds to university degree.
    \end{itemize}

The main goal of the analysis is to understand differences among ethnic groups, but  
we are also interested in assessing the effect of sex,  parity and gestational age
of the children on the development of obesity  \citep{tint2016abdominal}.
Sex, 
age and parity have been reported in the medical literature as associated to neonatal adiposity. 
Girls are
known to have greater adiposity than  boys even at birth \citep{simonetal:12,fieldsetal:09,PMID:15168110}.
Increasing parity is associated with increasing neonatal
adiposity in Asians as well as in Western populations \citep{joshietal:12,catalanoetal:95}.
Gestational age and postnatal age have also been shown to be
associated with increasing weight and adiposity \citep{simonetal:12,catalanoetal:95}.
Other important factors relating to the mother are  the results of the glucose tolerance test and pre-pregnacy body mass index, since metabolic diseases  
are heritable, though they do not necessarily lead to  obesity \citep{CDC}; see also, for instance,   \cite{qasim2018origin}. Since obesity might also be related to  family nutritional habits, we include in the model \textit{education} as proxy for the family socioeconomic status. 
 
  In the next subsection we present an exploratory data analysis (EDA), which will drive the choice of
interactions between the covariates  described above.

\subsection{Exploratory data analysis}
\label{sec:EDA}

The three main ethnic groups in Singapore are Chinese, Malay and Indian. Their sample frequencies in the dataset,  56\%, 26\% and 18\%, respectively, are consistent with  the overall distribution in the population. 
  
\begin{figure}[ht]
\centering
\includegraphics[width=0.6\textwidth]{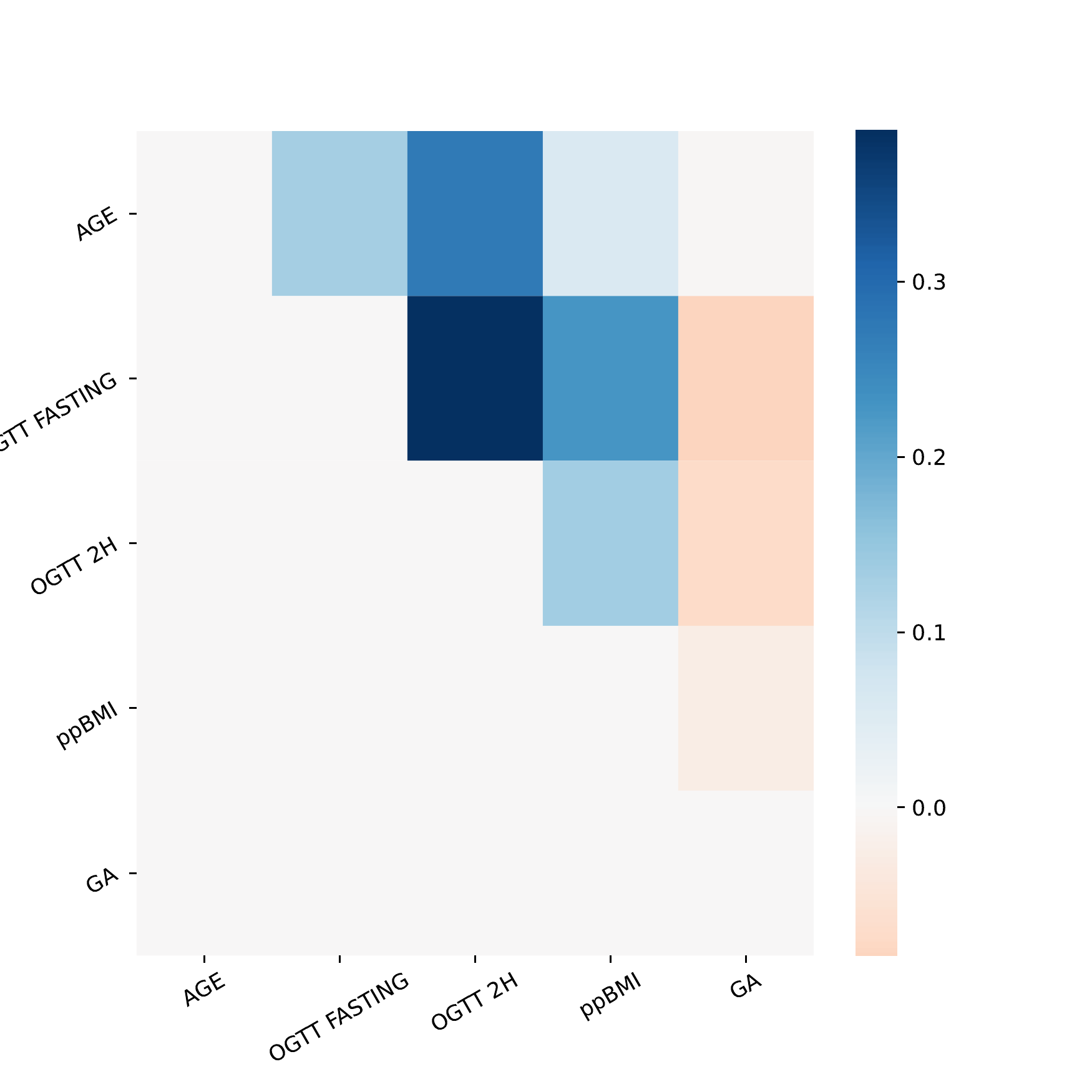}
\caption{Sample correlation between numerical covariates in Section~\ref{sec:data_description}}
\label{fig:corr_numericals}
\end{figure}

In 
Figure~\ref{fig:corr_numericals}
we plot the sample correlation of the numerical covariates.  We find that the largest correlation (equal to 0.42) is between \textit{OGTT fasting} and \textit{OGTT 2h}. 

\begin{figure}[!ht]
\centering
\includegraphics[height=0.7\textheight]{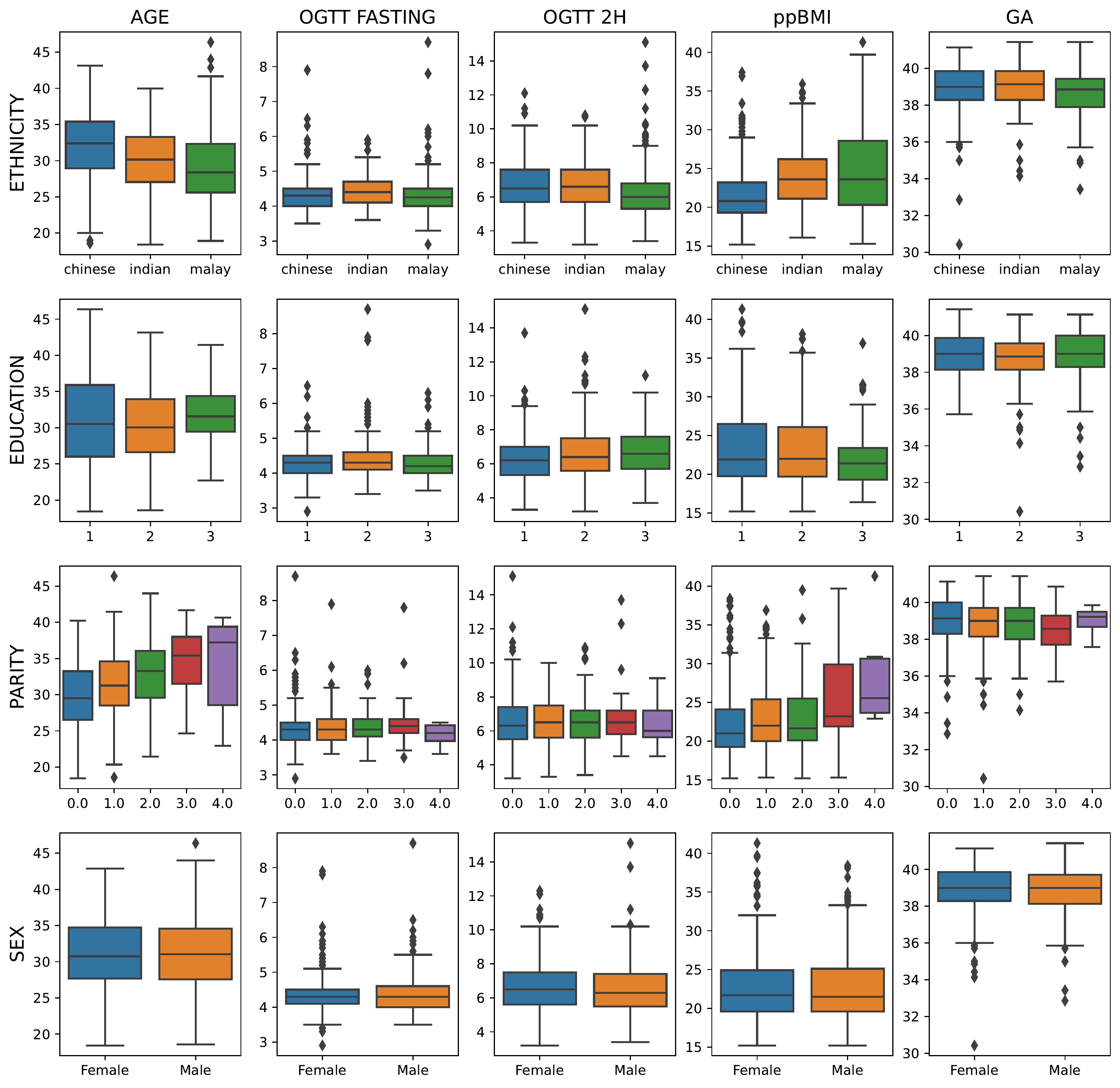}
\caption{Boxplots of numerical variables (by column) for each level of the categorical variables (by row).}
\label{fig:cat_vs_num}
\end{figure}

To understand   the relationship between categorical and continuous covariates, 
Figure~\ref{fig:cat_vs_num} shows histograms of each continuous  covariate, stratified  by each categorical covariate level. There appears to be a linear trend  between \textit{parity} and \textit{age}, which is to be expected,  and also
between \textit{parity} and \textit{ppBMI}. Additionally, the distribution of  mother's  is concentrated on  smaller values for Malay and Indian ethnicity, compared to Chinese women.  
No other association is detectable between categorical and continuous covariates.

 In Appendix~\ref{sec:fur_plots}  we show the unidimensional scatterplots  of the  responses (height and weight) at time $t=0,1,2$ versus the  continuous covariates, with the goal of identifying effect of  these covariates, which are time-homogeneous (recorded at baseline), on the responses (time-varying). For categorical covariates we plot  by boxplots of the responses stratified by level. See   Figure~\ref{fig:scatter_height}-\ref{fig:scatter_weight} in Appendix~\ref{sec:fur_plots}, which display a time-increasing response patterns,  though there does not seem to be a clear dependence of weight and height on the covariates.

\begin{figure}
    \centering
    \includegraphics[width=0.8\linewidth]{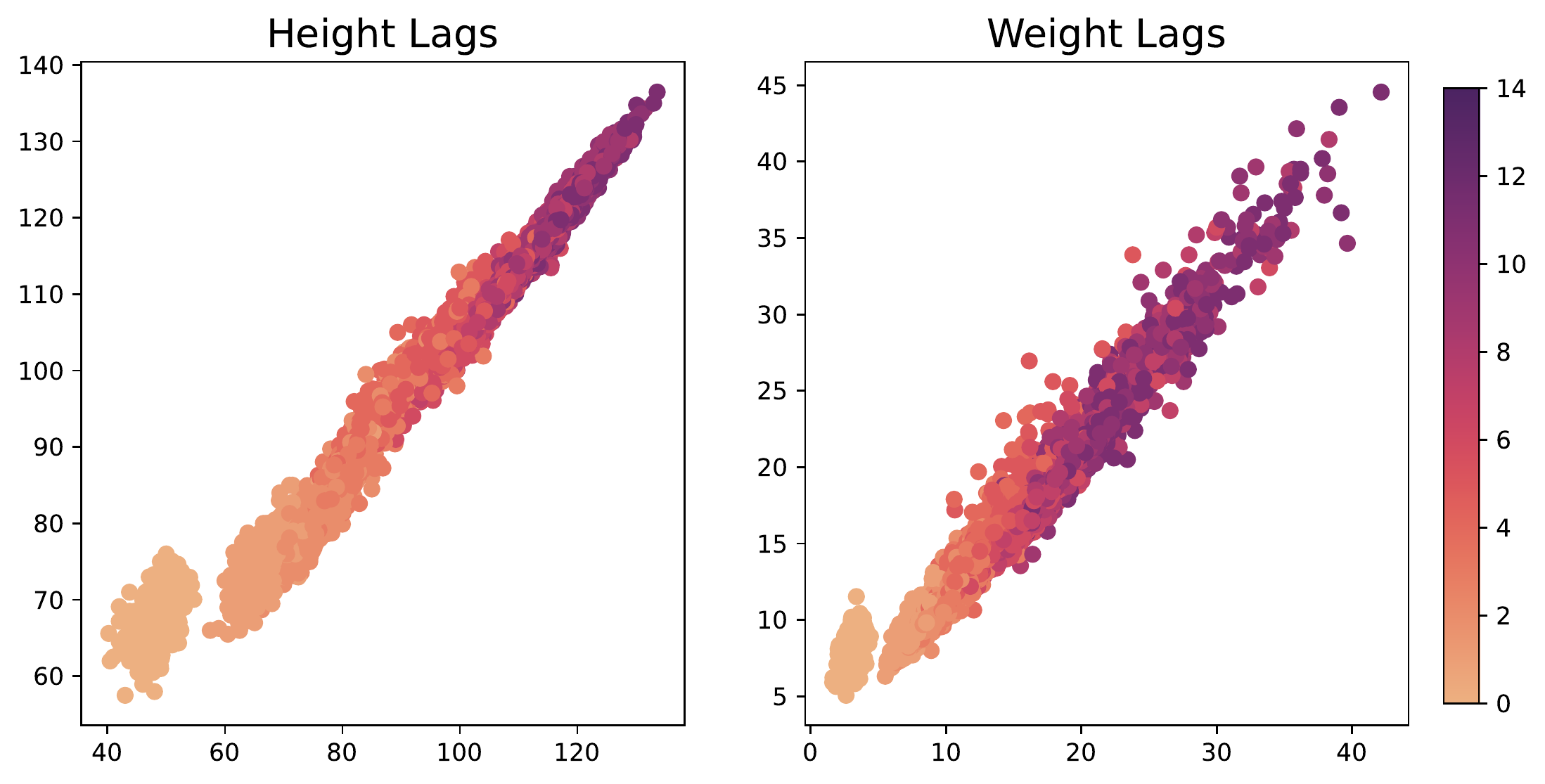}
    \caption{Scatterplots of Singapore children's height (left) and weight (right) at lag 1, i.e.  of the sample points $(y_{it},y_{i t+1})$, for $t=1,\ldots,T_i-1$ and $i=1,\ldots,N$ for response $y$; color corresponds to the age in the  colorbar}
    \label{fig:ts_lagged}
\end{figure}

 Figure~\ref{fig:ts_lagged} shows the scatterplots of the children's height (left) and weight (right) at lag 1, i.e. we plot sample  points $(y_{it},y_{i t+1})$ for all $t$ and all subject $i$ for both responses $y$. 
It is to identify two sub-groups in both plots, corresponding to newborns and infants (the group of datapoints on the left bottom corner) and older children. For the latter the autoregressive assumption is very clear, while for the infant group, as expected, the linearity assumption is not  strong, though it could be used as first approximation. 

As such, we propose a VAR model with lag 1 for the responses.
Moreover, we  include in the analysis  
the time-homogeneous covariates $\bm z_i$ and a function of  time, $x_{it} = \sqrt{t}$, as  time-varying covariate in the model, to account for a global
growth trend over time;  no other time-varying covariate is available in the dataset.  
We also consider  
interaction terms between (i)  the mother's highest education and age, and (ii) ethnicity
and sex of the child. 
Finally, denoting by $X:Y$ the interaction term between $X$ and $Y$, we include the following covariates in the model:
(1) an  intercept, (2) \textit{age}, (3) \textit{parity}, (4) \textit{OGTT fasting Pw26} (in what follows referred to as \textit{OGTT fasting}), (5) \textit{OGTT 2h Pw26} (in what follows referred to as \textit{OGTT 2h}), (6) \textit{ppBMI}, (7) \textit{GA}, (8) \textit{education$_{1}$:age} (9) \textit{education$_{2}$:age},(10) \textit{education$_{3}$:age}, (11) \textit{parity:age}, (12) \textit{Indian}, an indicator variable, equal to  if the mother is Indian and zero otherwise, (13) \textit{Malay} an indicator variable, equal to  if the mother is Malay and zero otherwise, (14) \textit{Male:Chinese}  indicator variable equal to 1 for a male child born to a Chinese mother, (15) \textit{Male:Indian} indicator variable equal to 1 for a male child born to an Indian mother and (16) \textit{Male:Malay}
indicator variable equal to 1 for a male child born to a Malay mother.  

The baseline category for the categorical covariates corresponds to  a female child born to a  Chinese mothers. 
As final pre-processing step, we standardize each numerical covariate at baseline  by subtracting their sample mean and dividing by the sample standard deviation. 

In summary, the  Child Growth dataset contains information on $N=766$ children, $k=2$ responses, $p=1$ time-dependent covariate (that is $\sqrt{t}$)   and a $q=14$-dimensional design matrix for time-homogeneous covariates (including intercepts, interactions and dummy variables to represent categorical covariates).

\section{The VAR model and the logit stick-breaking prior for the VAR parameters}
\label{sec:model}
Our motivating application requires the development of statistical methodology able to describe the evolution of a $k$-dimensional response vector $\bm Y_{it}$ for individuals $i$, $i=1,\ldots,N$ recorded  at discrete time points  $t$, $t=1,\ldots,T_i$, accounting  for  time-varying covariates  $\bm x_{it}$ and time-homogeneous covariates $\bm z_i$, measured at the baseline. 
Motivated  by the exploratory analysis in Section~\ref{sec:data}, we assume:
\begin{align}
\bm y_{it} = \Phi_i \bm y_{i t-1} + B \bm x_{it} + \Gamma \bm z_i + {\bm\varepsilon}_{it}, \quad {\bm\varepsilon}_{it} \iid \calN({\bm 0}, \Sigma) \quad t=1,\ldots,T_i, \ i=1,\ldots,N,
\label{eq:lik}
\end{align}
where $\Phi_i=[\Phi_{ijl}]$ is a $k\times k$ matrix of autoregression coefficients, $\bm x_{it}$ is a $p-$dimensional vector of time-varying covariates, $\bm z_i$   is a $q-$dimensional  vector of time-homogeneous covariates, $B=[b_{jl}]$  and $\Gamma=[\gamma_{jl}]$ are $k\times p$ and $k\times q$ matrices  of regression coefficients, respectively. For ease of explanation, we vectorize matrices $\Phi_i$, $B$ and $\Gamma$.  Specifically, denoting with  $(\cdot)^T$ the transpose of a column vector, we introduce the following notation
\begin{align*}
 \varphi_i &=(\Phi_{i11},\ldots,\Phi_{i1k},\Phi_{i21},\ldots,\Phi_{i2k},\ldots, \Phi_{ik1},\ldots,\Phi_{ikk})^T\\
\bm b &= (b_{11},\ldots,b_{1p},b_{21},\ldots,b_{2p},\ldots,b_{k1},\ldots,b_{kp})^T\\
\bm \gamma &= (\gamma_{11},\ldots,\gamma_{1q},\gamma_{21},\ldots,\gamma_{2q},\ldots,\gamma_{k1},\ldots,\gamma_{kq})^T,
\end{align*}
so that $\varphi_i$, $\bm b$ and $\bm \gamma$ are vectors with $k^2$, $k \times p$ and $k \times q$ elements (vectorization of the matrices $\Phi_i, B,\Gamma$, respectively).
We  assume  ${\bm y}_{i0}=\bm 0$, that is, conditionally to the remaining parameters, $\bm y_{i1}$ has a  Gaussian distribution with mean $B \bm x_{i1} + \Gamma \bm z_i$. Alternatively, we could consider the responses at baseline as exogenous.
Moreover, different initial distribution could be specified. 
We assume that a priori $({\Phi}_1,\ldots,{\Phi}_N)$, $\bm b$, $\bm\gamma$ and $\Sigma$ are independent.
As  random effect distribution we assume a Bayesian nonparametric prior which depends on the baseline covariates.  Specifically, we assume that 
\begin{equation}
	\Phi_i \mid z_i \ind \sum_{h=1}^H w_h (\bm z_i) \delta_{\Phi_{0h}} \quad i = 1, \dots, N. 
\label{eq:stick}
\end{equation}
and we impose a stick-breaking construction on the weights $w_h$. As such, equation \eqref{eq:stick} defines a truncated stick-breaking prior with $H$ support points 
$\{\Phi_{0h} \}$ and covariate-dependent weights summing to 1. 
Similarly to  \cite{rigon_durante_logit}, we assume that the weights are generated via a logit stick-breaking construction,
that is, 
$w_1(\bm z_i) = \nu_1 (\bm z_i)$, and $w_h(\bm z_i) = \nu_h(\bm z_i) \prod_{l=1}^{h-1} \left(1 - \nu_l(\bm z_i)\right)$ for $h=1, \ldots, H-1$, and $\nu_H(\bm z_i) = 1$.
The dependence on the covariates $\bm z_i$ is introduced by assuming a logistic model for $\nu_h(\bm z_i)$:
\begin{equation}\label{eq:logit}
\begin{aligned}
   \text{logit}(\nu_h (\bm z_i)) & = \bm z_i^T \bm\alpha_h, \quad h= 1, \dots, H-1  \\
   \bm
   \alpha_h & \iid \calN_q(\mu_\alpha, \Sigma_\alpha), \quad h=1, \dots H-1
\end{aligned}
\end{equation}
An equivalent formulation of  \eqref{eq:stick} can be obtained by introducing auxiliary variables $c_i$'s (usually referred to as cluster allocation indicators) such that 
\[	
	c_i \mid z_i, \bm \alpha \sim \mbox{Categorical}\left( \{1, \ldots, H\}; \bm w(\bm z_i) \right),
\]
and letting $\Phi_i = \Phi_{0 c_i}$. 
The introduction of the $c_i$'s allows us to make a fundamental distinction between mixture components and clusters. 
In the following, we refer to any of the $\Phi_{0h}$'s as a \emph{component}, while we call a \emph{cluster} of observations   a (nonempty) set  $\{i : c_i = h \}$; see, for instance, \cite{argiento2019infinity}. 
The marginal prior \eqref{eq:stick} - \eqref{eq:logit} is represented by a finite, though large number of parameters, 
and can be regarded as the truncation of a dependent Bayesian nonparametric prior.  

We complete the prior specification with  the marginal parametric prior distributions of 
$\bm b$, $\bm\gamma$ and $\Sigma$:
\begin{equation}\label{eq:prior_parametric}
 \bm b \sim \calN_{kp}( \bm 0, \Sigma_B), \qquad
  \bm \gamma \sim \calN_{kq}( \bm 0, \Sigma_{\Gamma}), \qquad
	\Sigma^{-1} \sim \mathcal W(\Sigma_0, \nu),
\end{equation}
where $\mathcal W(\Sigma_0, \nu)$ denotes the Wishart distribution
with expectation equal to $\nu \Sigma_0$ for $\nu > p -1$.

To obtain more robust inference, we assume a hierarchical prior for the $\varphi_{0h}$'s:
\begin{flalign}
\varphi_{0h} | \varphi_{00}, V_0 & \iid \calN_{k^2}(\varphi_{00}, V_0),  \qquad h=1,\ldots, H
\label{eq:atoms_prior_I}\\
\varphi_{00}, V_0 | \varphi_{000}, \lambda, V_{00}, \tau_0 & \sim \mathcal{NIW}(\varphi_{000}, \lambda, V_{00}, \tau_0).
\label{eq:atoms_prior_II}
\end{flalign}
In \eqref{eq:atoms_prior_II}, $\mathcal{NIW}(\varphi_{000}, \lambda, V_{00}, \tau_0)$ denotes  the normal-Inverse Wishart distribution,
 i.e. $V_0  \sim \mathcal{IW}(\tau_0, V_{00})$ and $\varphi_{00} \mid V_0 \sim \mathcal N(\varphi_{000}, \lambda^{-1} V_0)$, where $\mathcal{IW}(\tau_0, V_{00})$ denotes the inverse-Wishart distribution defined over the space of $k^2 \times k^2$ symmetric and positive definite matrices with mean $V_0 / (\tau_0 - k^2 - 1)$.

Posterior inference is performed through a Gibbs sampler algorithm, as detailed in Appendix~\ref{sec:gibbs}. However, it is worth noting that the full-conditional of the weights parameters $\{\bm\alpha_h\}$ in Equation \eqref{eq:logit} can be derived in closed-form with the introduction of  auxiliary variables, using results in \cite{polson_etal_jasa_2013} and \cite{rigon_durante_logit}.
The full conditional distributions of $\bm b$ and $\bm \gamma$ are derived as in a standard multivariate 
Bayesian linear regression models.
The full conditionals of the atoms $\{\Phi_{0h}\}$ in the stick-breaking prior \eqref{eq:stick} are given in the blocked Gibbs sampling of \cite{ishwaran_2001_gibbs}.

The code has been implemented in \texttt{C++} and linked to \texttt{Python} via \texttt{pybind11} \citep{pybind11}.

\section{Simulation study}\label{sec:simulation}

We now present a simulation study to compare the performance of the proposed `approach in 
\eqref{eq:stick}-\eqref{eq:logit} versus a similar model but assuming  the $\Phi_i$'s to be generated as independent and identically distributed from a Dirichlet Process \citep[DP,][]{Ferguson:73} which is arguably the most popular Bayesian nonparametric prior. \citep[see, e.g.][]{MuQuJaHa:15}. 

We consider three different simulation  scenarios.
In  scenarios (I) and (II) the responses are simulated from \eqref{eq:lik}, while in  scenario (III) we simulate each $\varepsilon_{itj}$ in ${\bm\varepsilon}_{it} = (\varepsilon_{it1}, \ldots, \varepsilon_{itk})$ from a student-t distribution with mean 0 and 5 degrees of freedom, so that our model is then misspecified.
Of course, other kind of misspecifications are possible, for instance, we could generate data from an  autoregressive process with a larger lag, but this would lead to much poorer results for any model with our likelihood. 
For all  scenarios, we simulate $N=300$ independent trajectories  $\bm y_i = (\bm y_{i1}, \ldots, \bm y_{iT_i})$, with $T_i=10$ for all $i$,
assuming each $\bm y_{it}$ to be a three-dimensional vector (i.e., $k=3$). Moreover, we always set $B=\bm 0$, $\Gamma = 0$ in the data generating process.

\begin{figure}
    \centering
    \includegraphics[width=0.6\linewidth]{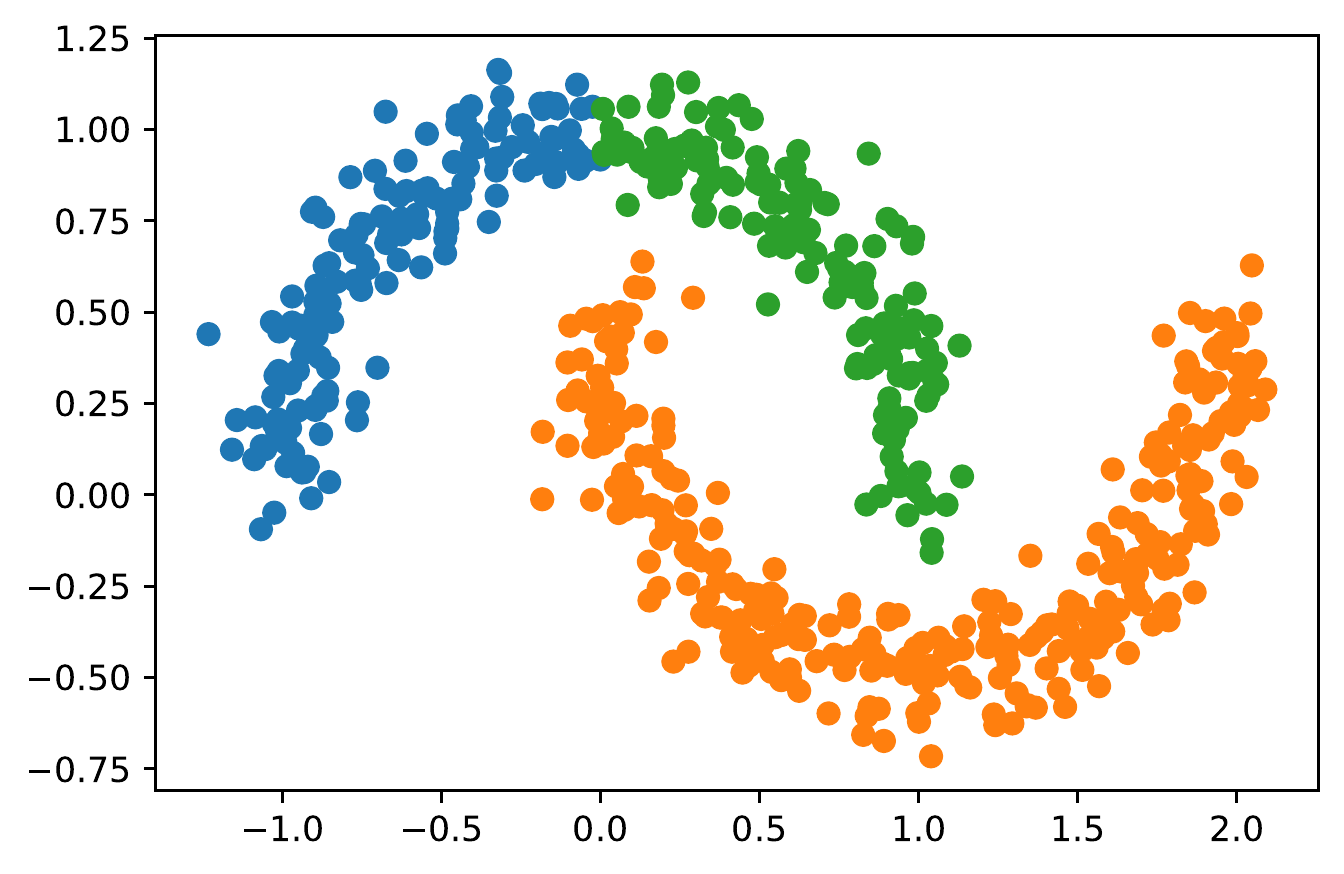}
    \caption{Fixed time covariates for  scenario (II)}
    \label{fig:s2_fixcov}
\end{figure}

 In all the scenarios, for each item  $i$, we simulate $\Phi_i$ from a discrete mixture, $ \Phi_i \iid \sum_{j=1}^3 \pi_j \delta_{\bar{\phi}_j}$, where the $\bar{\phi}_j$'s are given  in  \eqref{eq:simu_phimat} (see here below).
Then, conditionally to $\Phi_i$ we generate the time-homogeneous  covariate vector $\bm z_i$. 
In scenarios (I) and (III) the weights $(\pi_1, \pi_2, \pi_3)$ are set equal to $(0.5, 0.5, 0)$, $\bm z_i \mid \Phi_i = \bar \phi_{1} \sim \mathcal N_2((-3, -3),  I_2)$ and $\bm z_i \mid \Phi_i = \bar \phi_{2} \sim \mathcal N_2((3, 3), I_2)$ 
\begin{equation}
\label{eq:simu_phimat}
    \bar\phi_{1} = \begin{bmatrix} 1.1, 0.0, 0.0 \\ 0.0, 1.1, 0.0 \\ 0.0, 0.0, 1.0 \end{bmatrix} , \qquad \bar\phi_{2} = \begin{bmatrix} 1.1, -0,1, 0.0 \\ -0.1, 1.1, -0.1 \\ 0.0, 0.0, 0.9 \end{bmatrix}, \qquad \bar\phi_{3} = \begin{bmatrix} 0.9, -0,1, 0.0 \\ -0.1, 1.1, -0.1 \\ -0.1, 0.0, 1.5 \end{bmatrix},
\end{equation}
while in scenario (II) the weights are $(0.25, 0.25, 0.5)$ and the simulated time-homogeneous covariates are reported in Figure~\ref{fig:s2_fixcov}.
Observe that while in  scenario  (I) and (III) the covariates in the different clusters are clearly separable, this is no longer the case in  scenario (II).
Finally, in scenarios (I) and (II)  we fix $\Sigma = 0.25 I$ in \eqref{eq:lik}, while in  scenario (III) the error terms are generated from a student-t distribution as previously explained.

In the simulations, we set the hyperparameters in \eqref{eq:atoms_prior_II}  as follows: 
$\Phi_{000}=0$, $\lambda=0.1$, $V_{00}=I_9$, $\tau_0 = 11$.
Moreover, we fix $\Sigma_0 = I_3/\nu$ and $\nu=5$ (see \eqref{eq:prior_parametric}), so that $\Sigma^{-1}$ has prior mean equal to $I_3$. 
For our model, we further assume $\mu_\alpha = \bm0$, $\Sigma_\alpha= I_9$, $\Sigma_B=I_2$ and $\Sigma_\Gamma=I_{18}$; see \eqref{eq:logit}-\eqref{eq:prior_parametric}.
For the alternative Dirichlet process prior on the $\Phi_i$'s, we consider the truncated stick-breaking approximation \citep{ishwaran_2001_gibbs},
with total mass parameter equal to $1$.
For both priors the number of atoms $H$ is set equal to 25.

We assess predictive performance of both models through out-of-sample  prediction 
and 
$l$-steps ahead in-sample   prediction
for observed samples.
In the first case (later referred to as OOS), for all scenarios, we generate a new test set  of size $300$ following the
same data generating process outlined above, while in the second experiment (INS)
we randomly pick $100$ of the $300$ trajectories   generated   and
\virgolette{truncate} them at $T=5$.
In the first setting, the goal is to predict the whole time trajectory given responses at time 1. 

We expect our model under prior \eqref{eq:stick}-\eqref{eq:logit} to perform much better than when $\Phi_i$ are iid from the
Dirichlet process, since our model can assign data to clusters based
on their time-homogeneous  covariates, while the DP prior does not.
In the second setting, the goal is to predict $l=5$ steps in the future, i.e. predict $\bm y_{i6}, \ldots, \bm y_{i 10}$, for the $100$  truncated trajectories. 
Observe that in this case, we condition on the cluster membership inferred through the MCMC simulation, so that the fixed time covariates are not used to assign trajectories to clusters. As such, we expect the DP prior to have better predictive performance than our model since the number of parameters is considerably smaller compared to our model.
Finally, we also consider the quality of the estimated random partition of the subjects/datapoints, by computing the Adjusted Rand Index \citep{hubert1985comparing} between the point estimate of
the partition, obtained by minimizing the Binder loss function with equal
missclassification cost \citep[see, e.g.,][]{lau2007bayesian}, 
based on the MCMC samples and the true partition
given by the data generating process.

\begin{table}
\centering
\begin{tabular}{c | c c | c c | c c}
& \multicolumn{2}{c|}{Scenario (I)} & \multicolumn{2}{c|}{Scenario (II)} & \multicolumn{2}{c}{Scenario (III)} \\\hline
& LSB & DP & LSB & DP & LSB & DP\\
\hline
OOS & $7.5 \pm 6.6$ & $65.41 \pm 58.8$ & $5.8 \pm 3.3$ & $41.9 \pm 23.5$ & $91 \pm 141$ & $623 \pm 1604$ \\
INS & $3.45 \pm 3.12 $ & $3.43 \pm 3.02$ & $3.9 \pm 8.4$ & $4.0 \pm 8.5$  & $60.7 \pm 114$ & $60.4 \pm 113$ \\
ARI & 1.0 & 1.0 & 0.98 & 0.9 & 1.0 & 1.0
\end{tabular}
\caption{
Simulated dataset: out-of-sample (OOS) and  in sample (INS) mean squared prediction errors
and Adjuster Rand Index (ARI) for our model (LSB) and the Dirichlet Process prior for $\Phi_i$'s parameters (DP).}
\label{tab:simu_lsb_dp}
\end{table}
Goodness-of-fit indices  shown in Table~\ref{tab:simu_lsb_dp} confirm our expectations for the  out-of-sample testing setting(OOS), that is the proposed approach (denoted in the table as LSB, \textit{logit stick-breaking}) outperforms the DP prior in terms of mean squared prediction. It is clear  that for the in-sample predictions both models  performs similarly.  Note that our model has a slightly better accuracy in terms of  clustering for setting (II). This is likely due to the fact that clustering estimation is based also on covariate information and not only on  response patterns. The posterior distribution from the DP model favours a larger number of clusters  to better approximate the heavy tails of the error's distribution. 

\begin{figure}
\centering
\includegraphics[width=\linewidth]{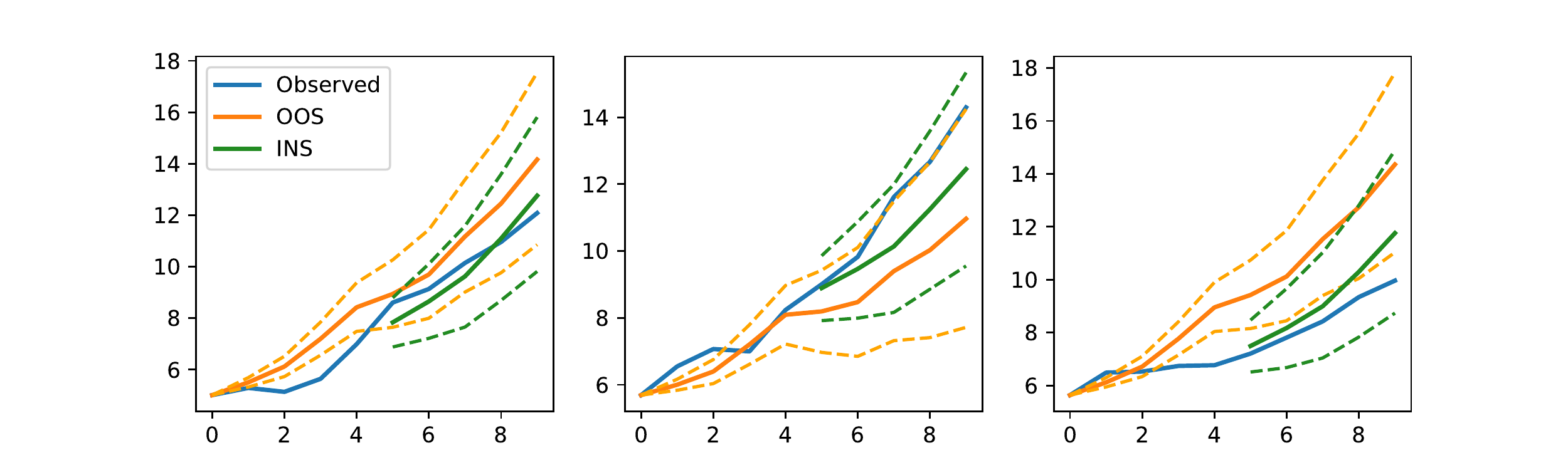}

\includegraphics[width=\linewidth]{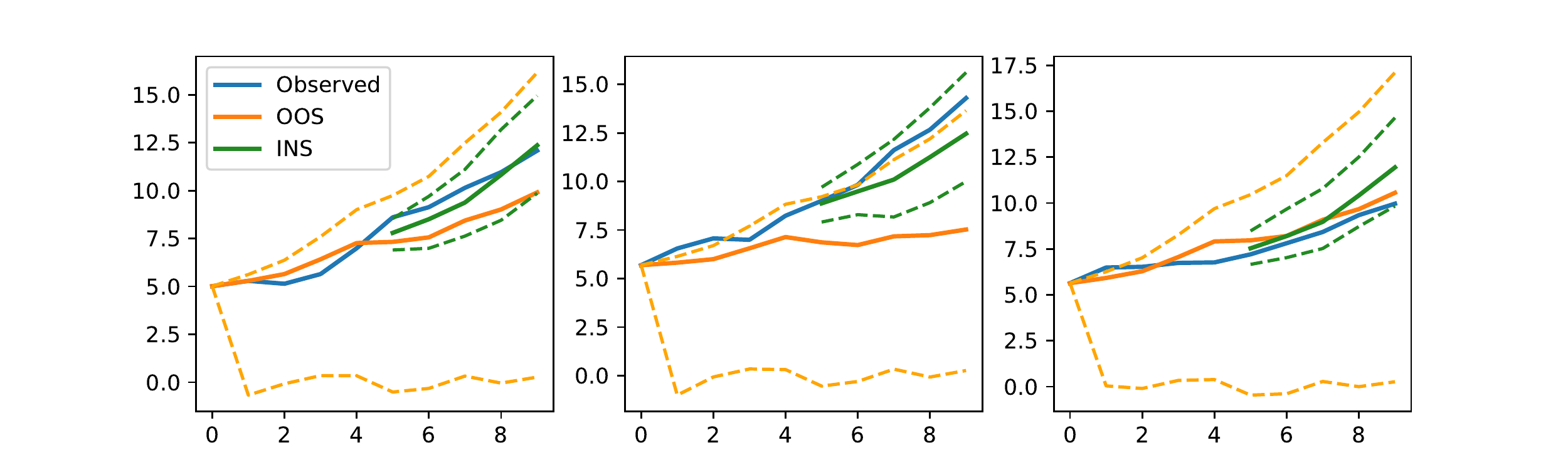}
\caption{Posterior predictive distributions for both  priors under comparison, considering both out-of-sample and in-sample predictions for scenario (I). 
We show predictive density estimates and credible intervals using our model (top row) and the DP prior (bottom row) for a new subject $i$.
In each panel, the solid blue lines denote the observed trajectory.
The OOS prediction (i.e. given $\bm z_i$ and $\bm y_{i1}$) is shown in orange, while the INS prediction (i.e. given $\bm y_{i5}$ and the cluster label $c_i$) is shown in green.
Solid lines correspond to the median for each time while dashed lines correspond to $95\%$ credible bands of the predictive distributions.}
\label{fig:simu_lsb_dp}
\end{figure}

Figure~\ref{fig:simu_lsb_dp} shows posterior predictive distributions for both  priors under comparison, considering both out-of-sample and in-sample predictions for scenario (I). 
We can see that in the OOS case the credible bands for the DP prior are very wide, while those under
our model are much narrower. 
Further, in the INS case, both models display better predictive performance and
narrower credible bands.

\begin{figure}
    \centering
    \includegraphics[width=0.7\linewidth]{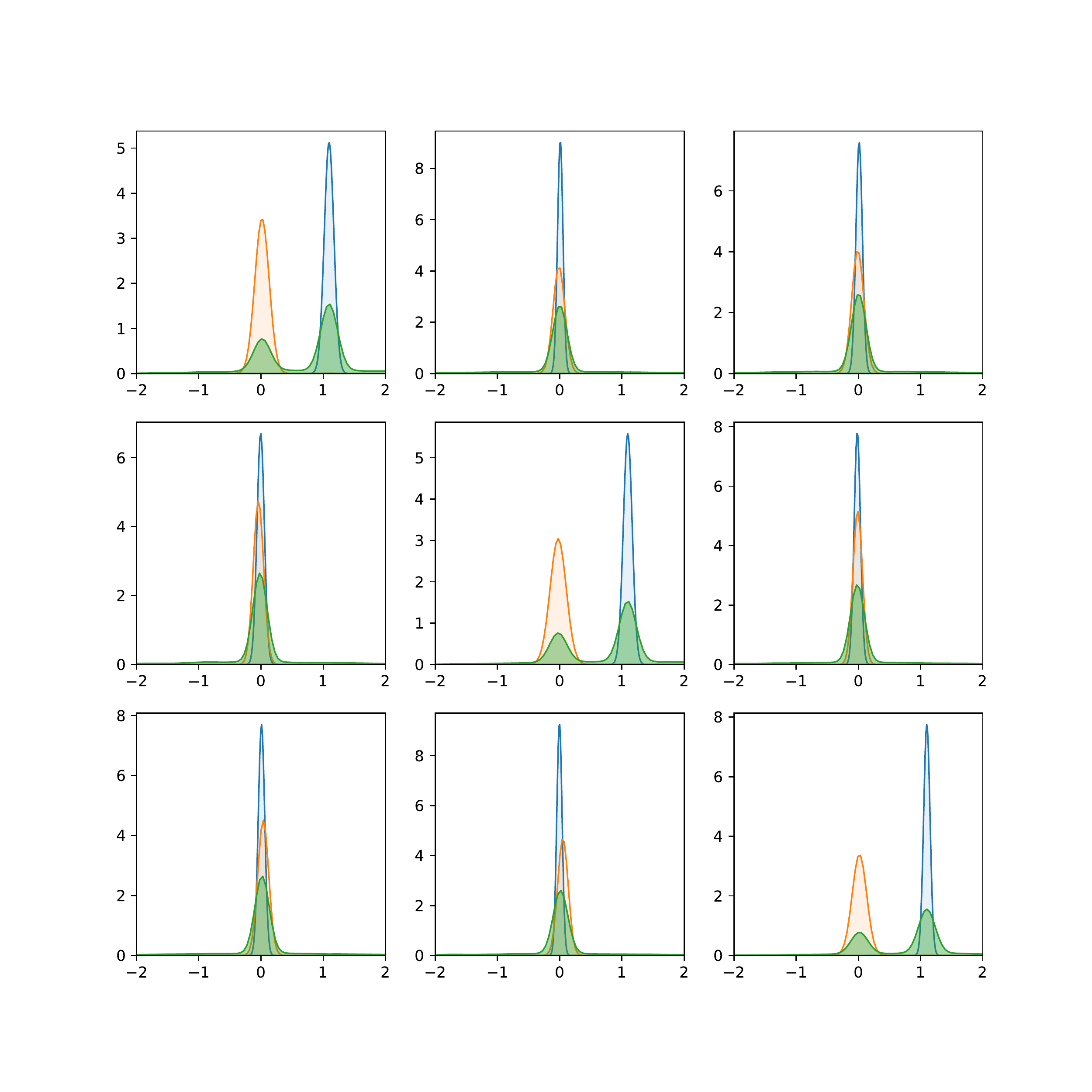}
    \caption{Predictive distributions of $\Phi_i^{new}$ corresponding to three subjects with fixed-time covariates equal to $(-3, 0)$ (blue) $(3, 0)$ (orange) and $(0, 0)$ (green), respectively.}
    \label{fig:pred_phi}
\end{figure}

Finally, we simulate a  new dataset under scenario (I), but fixing $\bar\phi_2 = \bm 0$ so that the corresponding trajectories are well separated; we note that there is no substantial difference in posterior inference. 
Figure~\ref{fig:pred_phi} reports kernel density estimates from the MCMC sample of the 
predictive distribution of $\Phi_i$  for  scenario (I), for three new observations with time-homogeneous  covariates equal to $(-3, 0)$, $(3, 0)$ (which coincide with the means of the first and the second group of simulated data) and $(0, 0)$ respectively.
Note that the predictive distribution associated to  covariate vector  equal to $(0,0)$ (reported in green in Figure~\ref{fig:pred_phi})
is bimodal, giving almost equal mass to values near
$\bar\phi_1$ and $\bar\phi_2$. 

This simulation study shows that the proposed models based on a covariate dependent prior outperforms non-dependent alternative prior in terms of prediction.  Moreover, also in terms of clustering structure recovery, 
the covariate dependent prior gives better estimates in case of heavy tail data.

\section{ Child Growth data}\label{sec:datanalysis}
In this section we present posterior results for the Child Growth dataset, detailing prior specification   (Section~\ref{sec:prior_elicitation}) and 
 inference in Section~\ref{sec:post_inf}/   

\subsection{Prior elicitation}
\label{sec:prior_elicitation}

Given the complexity of the model and the high-dimensionality of the dataset, prior elicitation needs to be carefully considered.
Preliminary analysis shows that
 that when the variances of the $\alpha_h$'s  (see \eqref{eq:logit}) or of the atoms $\Phi_{0h}$'s  (see \eqref{eq:atoms_prior_I}) in the logit stick-breaking are large, then 
all the observations tend to be assigned to the same component.
Moreover, the missing data simulation step has a strong impact on posterior inference.
In particular, when using  the vague prior described above, in   the initial iterations of  the MCMC algorithm, typically large missing values were imputed (e.g. $10^5$) 
since both $\Sigma$ and $\{\Phi_{0h}\}$ would take on unusually large values.
Consequently, sampled values for all the other parameters are affected, leading to a poor fit. 
Hence the use of
 an uninformative  prior is not advisable, causing poor mixing and slow convergence of the chain. Moreover, this is a common situation in complex hierarchical models when non-informative priors are adopted in lower levels.

\begin{figure}[ht]
\centering
\includegraphics[width=0.8\linewidth]{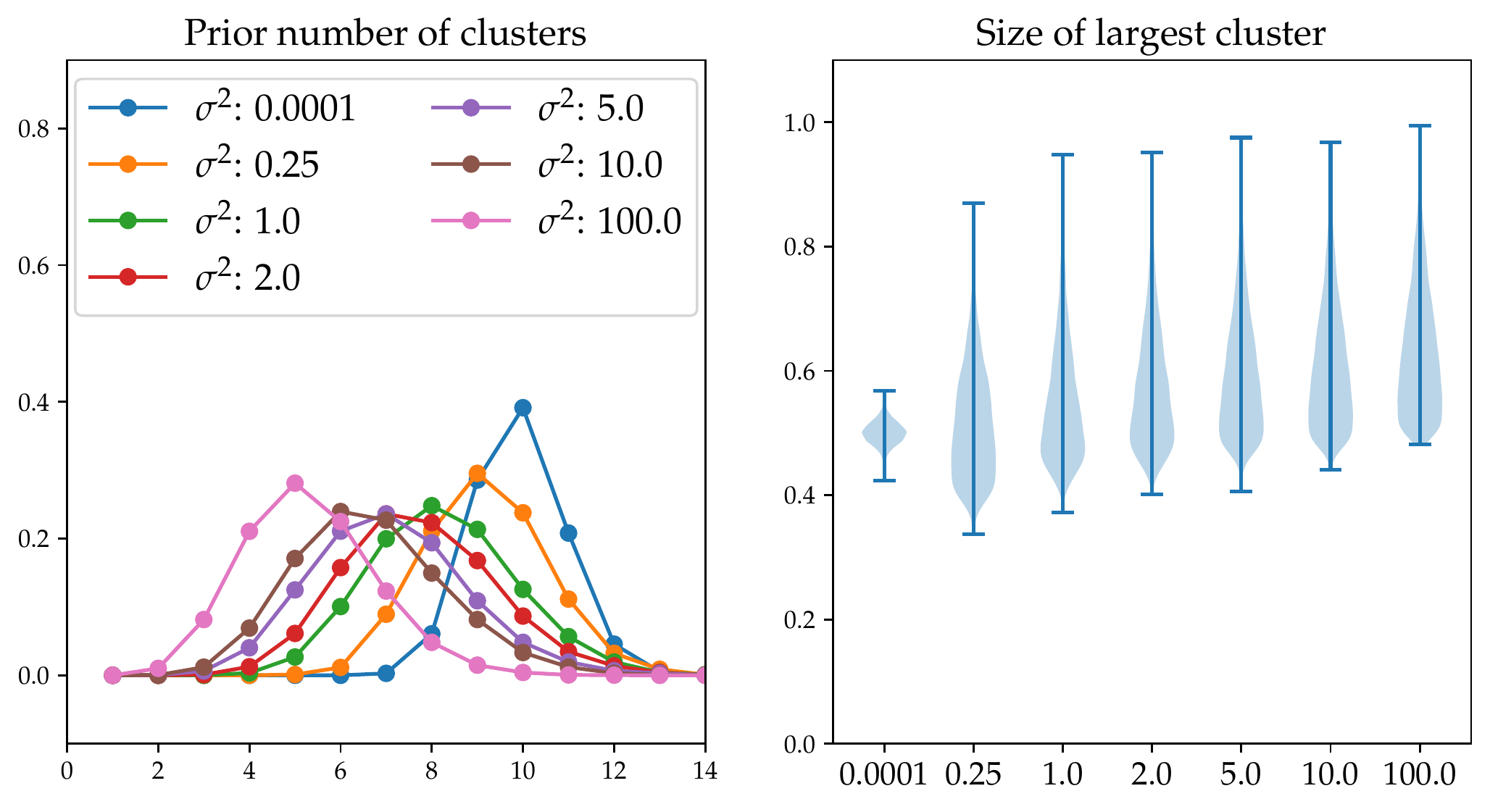}
\caption{Prior distribution of the number of clusters (left panel) and of the size of the largest cluster  as percentage of the whole dataset (right panel),  for different values of $\sigma_\alpha$.}
\label{fig:prior_simu}
\end{figure}

As such, we opt for informative priors.  
To set the hyperparameters in the hierarchical marginal prior in \eqref{eq:atoms_prior_I}-\eqref{eq:atoms_prior_II}, we first obtain the maximum likelihood estimator from a vector autoregressive model:
\begin{equation}\label{eq:ml}
    \bm y_{it} \mid \bm y_{i t-1} \sim \calN(\Phi \bm y_{it-1}, \Sigma), \qquad t=1, \ldots T - 1, \, i=1, \ldots, N 
\end{equation}
which corresponds to \eqref{eq:lik} when $B$ and $\Gamma$ are set to zero (their prior expected value) and $H=1$. 
We fit \eqref{eq:ml} using only subjects with no missing responses.
Let $\widehat{\Phi}$, $\widehat{\Sigma}$ denote the maximum likelihood estimator for $\Phi$ and $\Sigma$ respectively.
We fix $\Phi_{000} = \widehat{\Phi}$, $\lambda = 1$, and select $(V_{00}, \tau)$ in \eqref{eq:atoms_prior_II} so that $\mathbb{E}[V_0] = I$ and $\text{Var}[\{V_0\}_{ii}] = 1.5$.
Similarly, we fix $\Sigma_{0}$ and $\nu$ in \eqref{eq:prior_parametric} so that $\mathbb{E}[\Sigma] = \widehat{\Sigma}$ and $\text{Var}[\{\Sigma_{ii}\}] = 10$. 
The variance hyperparameter $\Sigma_\alpha$ in \eqref{eq:logit} also has an important effect on posterior inference.
To set this quantity, we look at the  prior distribution of the number of clusters (i.e. \textit{occupied components}) and of the size of the largest cluster.  To this end, we perform Monte Carlo simulations. 
Specifically, we fix the number of components $H$ in the stick-breaking prior equal to $50$,
set $\Sigma_\alpha = \sigma^2_\alpha I$, and simulate $\alpha_1, \ldots, \alpha_{H-1}$ from \eqref{eq:logit} with $\mu_\alpha = \bm 0$. Then, for each of the $N=766$ subjects, we compute the associated weights $\bm w(\bm z_i)$ from the logit stick-breaking process, using observed covariates $\bm z_i$, and allocate each subject to one of the $H$ components with probability given by the weights $\bm w(\bm z_i)$ The above procedure is repeated independently for $M=10,000$ iterations and we record the number of clusters and the size of the largest cluster.
Figure~\ref{fig:prior_simu} shows the distributions obtained from the Monte Carlo simulation.
As $\sigma_\alpha^2$ increases, the number of clusters shrinks to 1 and the size of the largest cluster increases accordingly. Hence, we fix $\sigma_\alpha^2 = 5$ so that a priori we should expect approximately $4-7$ clusters. 
Finally, we assume $\mu_\alpha = \bm0$, $\Sigma_B=I_2$ and $\Sigma_\Gamma=I_{18}$ (see \eqref{eq:prior_parametric}); recall that all continuous covariates are standardized. 

\subsection{Posterior inference results}
\label{sec:post_inf}

We apply the model described in Section~\ref{sec:model} to the Child growth dataset with hyperparameters set as in Section~\ref{sec:prior_elicitation}. 
We run the MCMC algorithm  for $100,000$ iterations, discarding the first 50,000  as burn-in and thinning  every 10 iterations, obtaining a final sample size of 5,000 iterations.

\begin{figure}[!ht]
\centering
\includegraphics[width=0.6\linewidth]{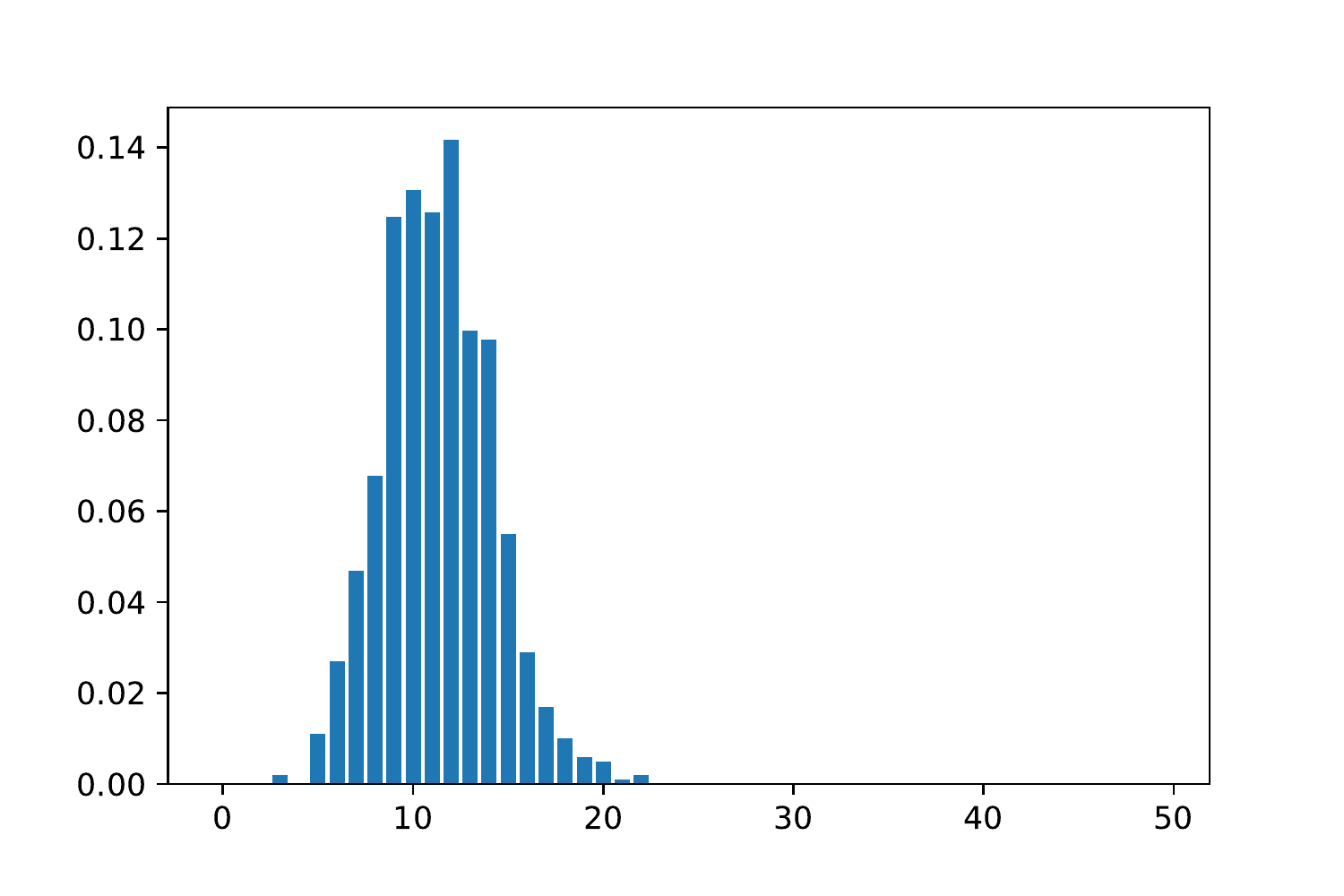}
\caption{Child Growth dataset: posterior distribution of the number of clusters.}
\label{fig:used_components}
\end{figure}

\begin{figure}[ht]
\centering
\includegraphics[width=0.75\linewidth]{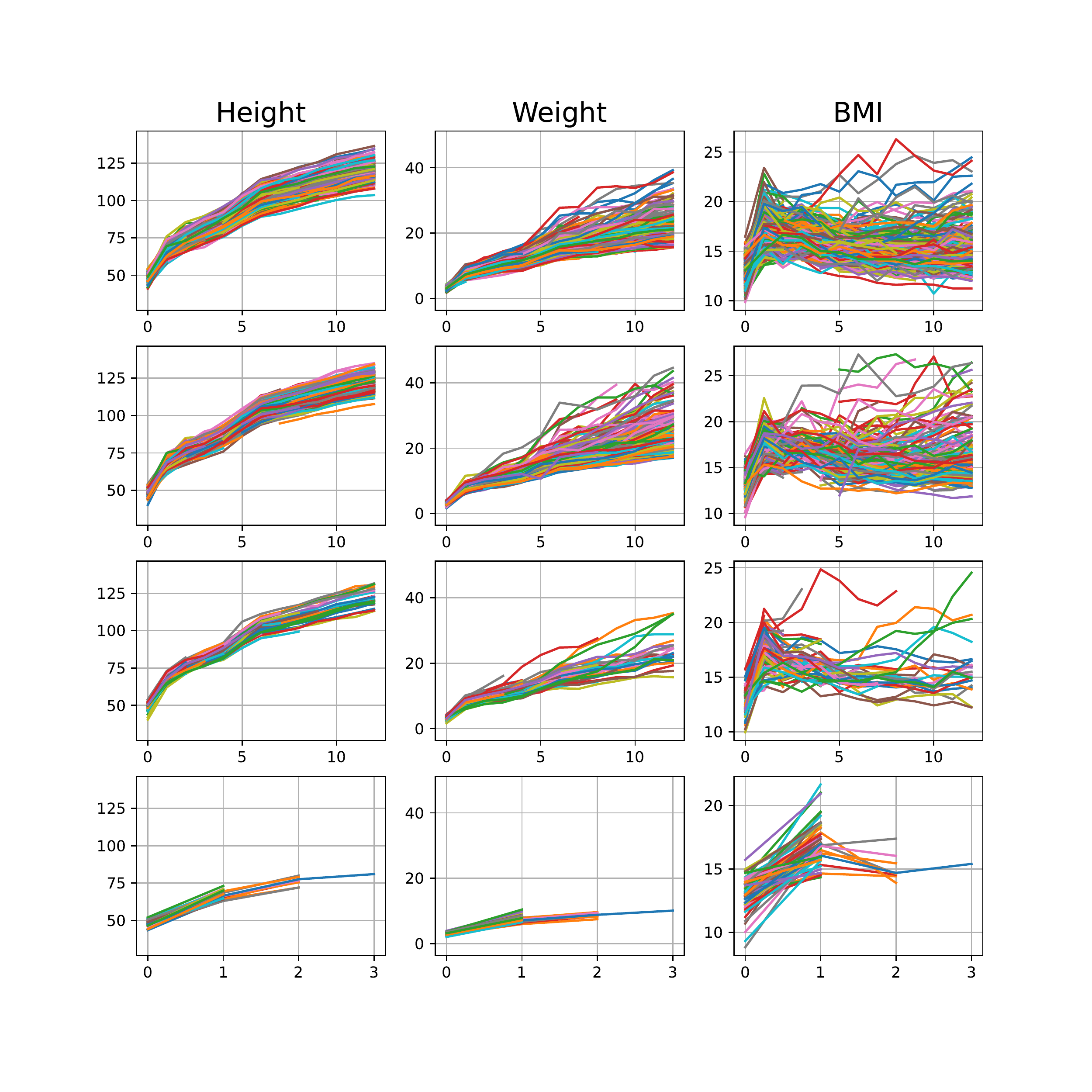}
\caption{Subject trajectories of height (first column), weight (second column) and BMI (third column) by estimated cluster (by row). The figure reports only the four largest clusters out of the seven estimated.}
\label{fig:clusters}
\end{figure}

Figure~\ref{fig:used_components} shows the posterior distribution of the number of clusters, i.e. of \textit{occupied} parametric components, that is clearly centered around 10-12 clusters. However, interpreting these as the \virgolette{number of distinct profiles} in the $\bm y$'s  may be misleading.  Recall that we have specified a covariate-dependent prior for the
random partition of patients.
Indeed, some clusters can be essentially identical when looking at the response trajectories 
but different when looking at the covariates. 
As a point estimate of the latent partition,  we choose the one that   minimises  the Binder loss function under equal misspecification costs \citep{binder1978bayesian}. The estimated partition consists of seven clusters, of which only four contain least 15 observations. In  
Figure~\ref{fig:clusters} we display the response trajectories clustered according to the estimated partition.
Note that the fourth cluster (bottom row) consists of subjects with at most three visits, except for one single subject with four visits. For this reason, we do not discuss this cluster. Figure~\ref{fig:clusters} shows the time trajectories for patients' height (first column),  weight (second column) and BMI.
The third row in Figure~\ref{fig:clusters} shows that this cluster contains children with lower weight, and consequently
 lower BMI  than the other two clusters. 

\begin{figure}[ht]
    \centering
    \begin{subfigure}{0.3\linewidth}
        \includegraphics[width=\linewidth]{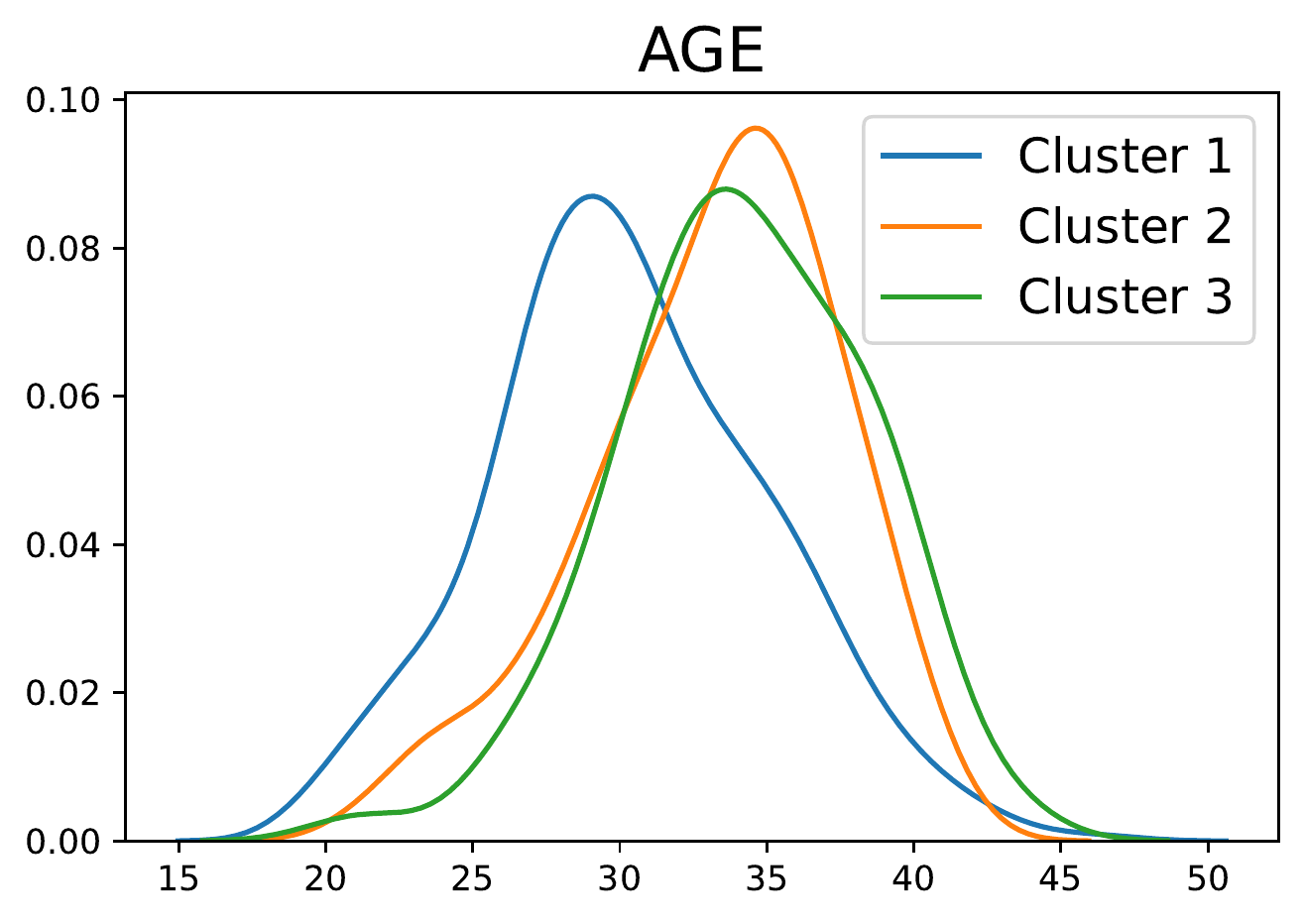}
        \caption{0.00, 0.00, 0.61}
    \end{subfigure}%
    \begin{subfigure}{0.3\linewidth}
        \includegraphics[width=\linewidth]{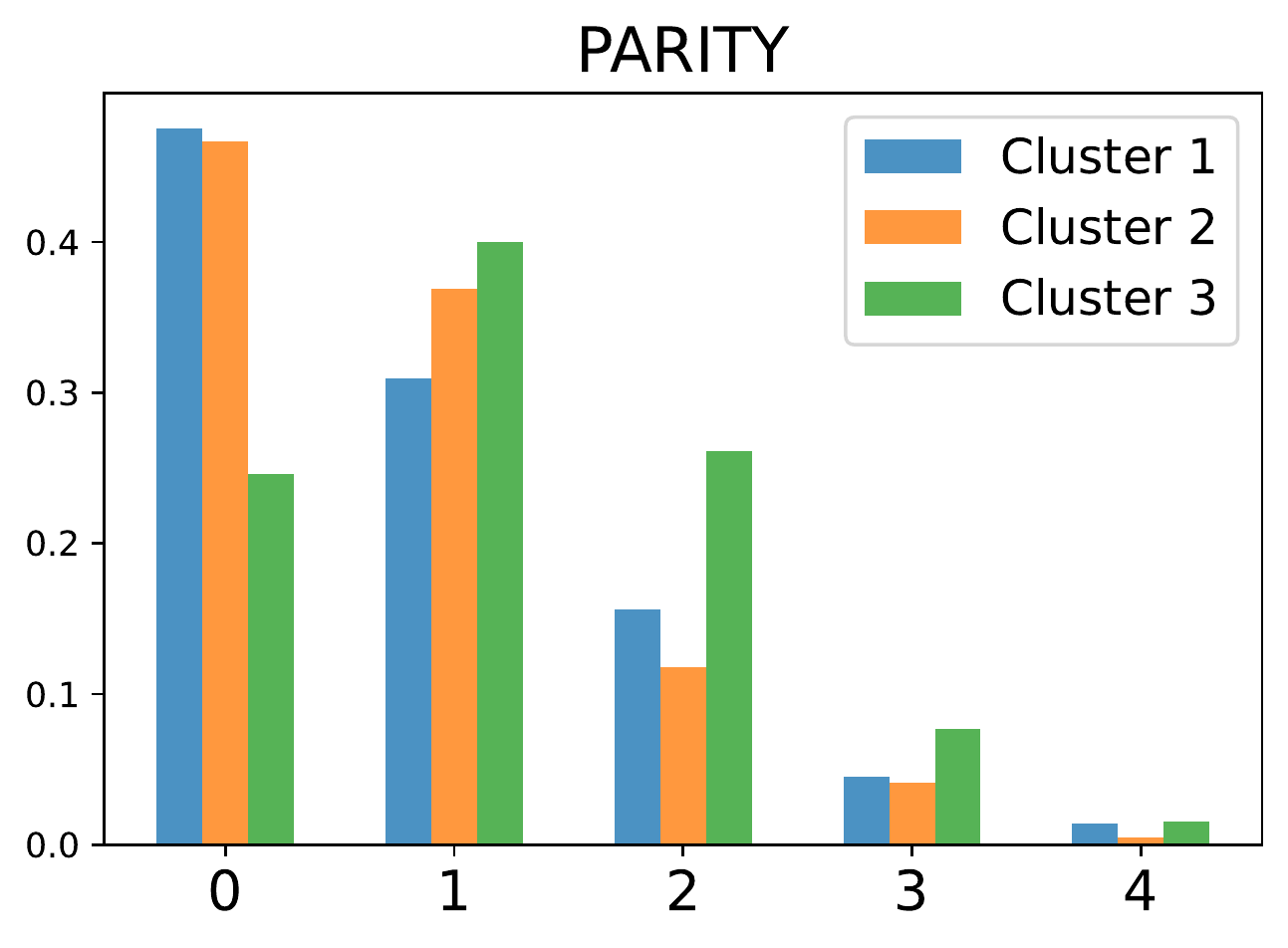}
        \caption{0.02, 0.00, 0.0}
    \end{subfigure}%
    \begin{subfigure}{0.3\linewidth}
        \includegraphics[width=\linewidth]{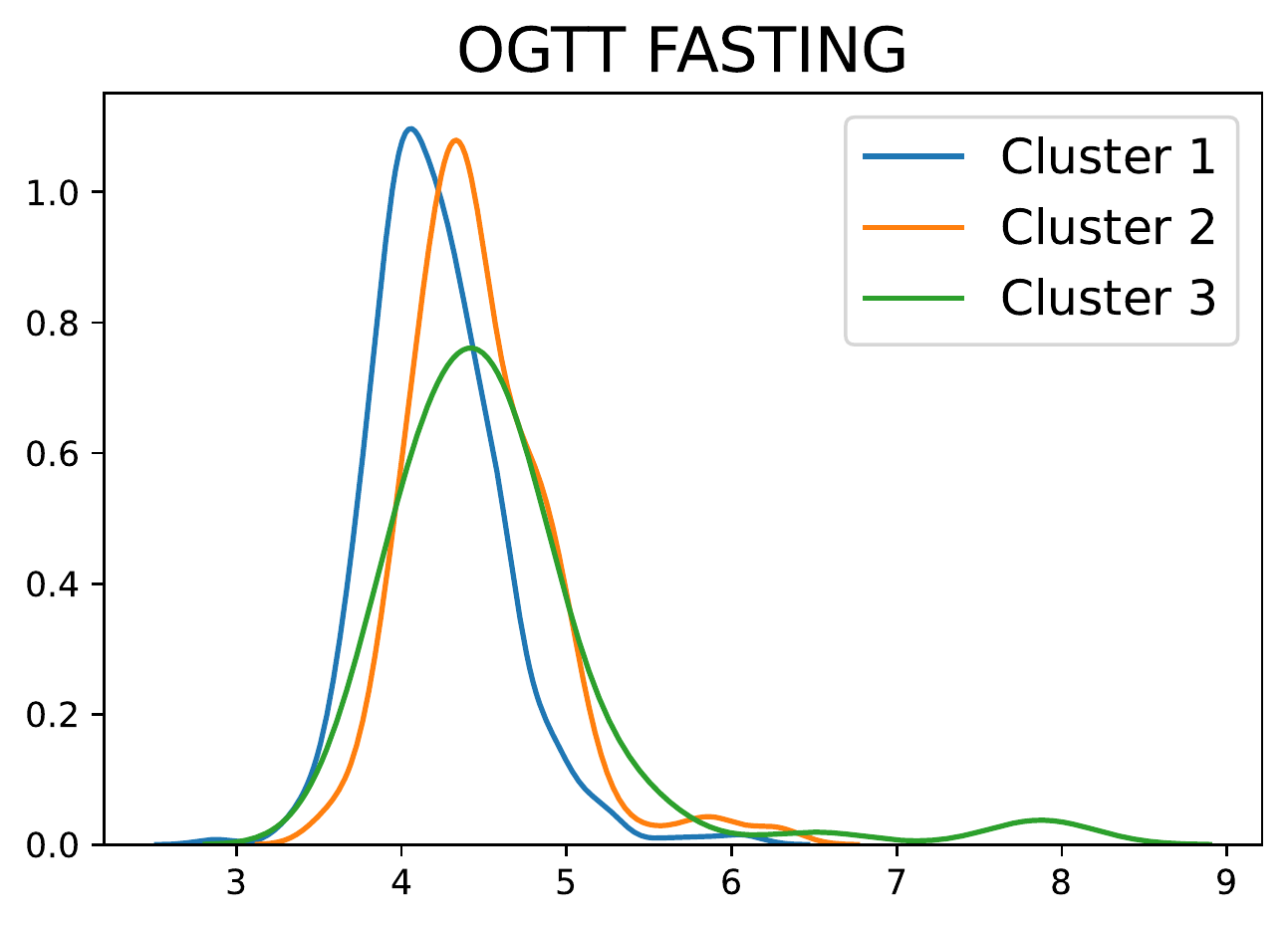}
        \caption{6e-10, 3e-5, 0.66}
    \end{subfigure}

    \begin{subfigure}{0.3\linewidth}
        \includegraphics[width=\linewidth]{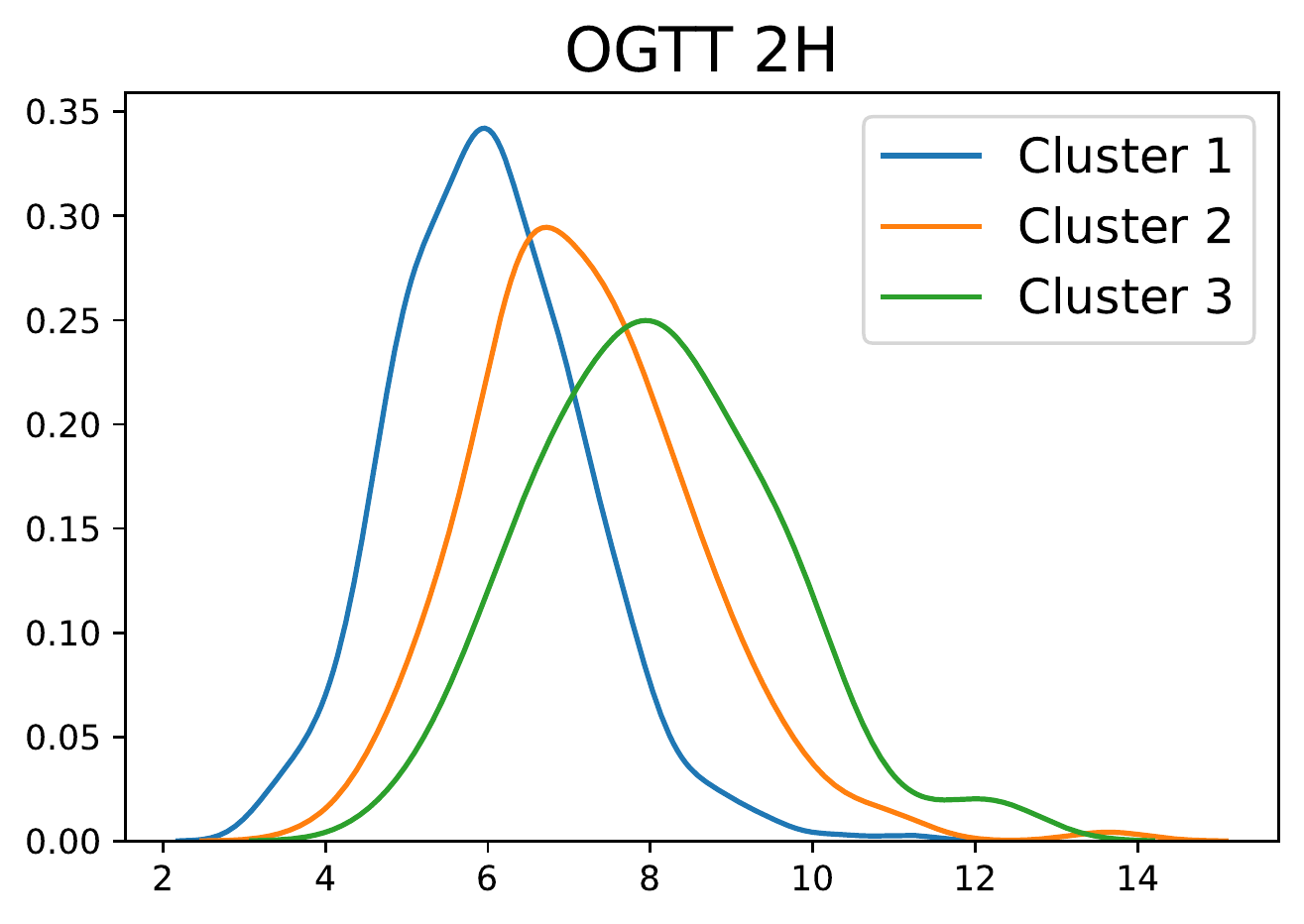}
        \caption{0.00, 0.00, 0.00}
    \end{subfigure}%
    \begin{subfigure}{0.3\linewidth}
        \includegraphics[width=\linewidth]{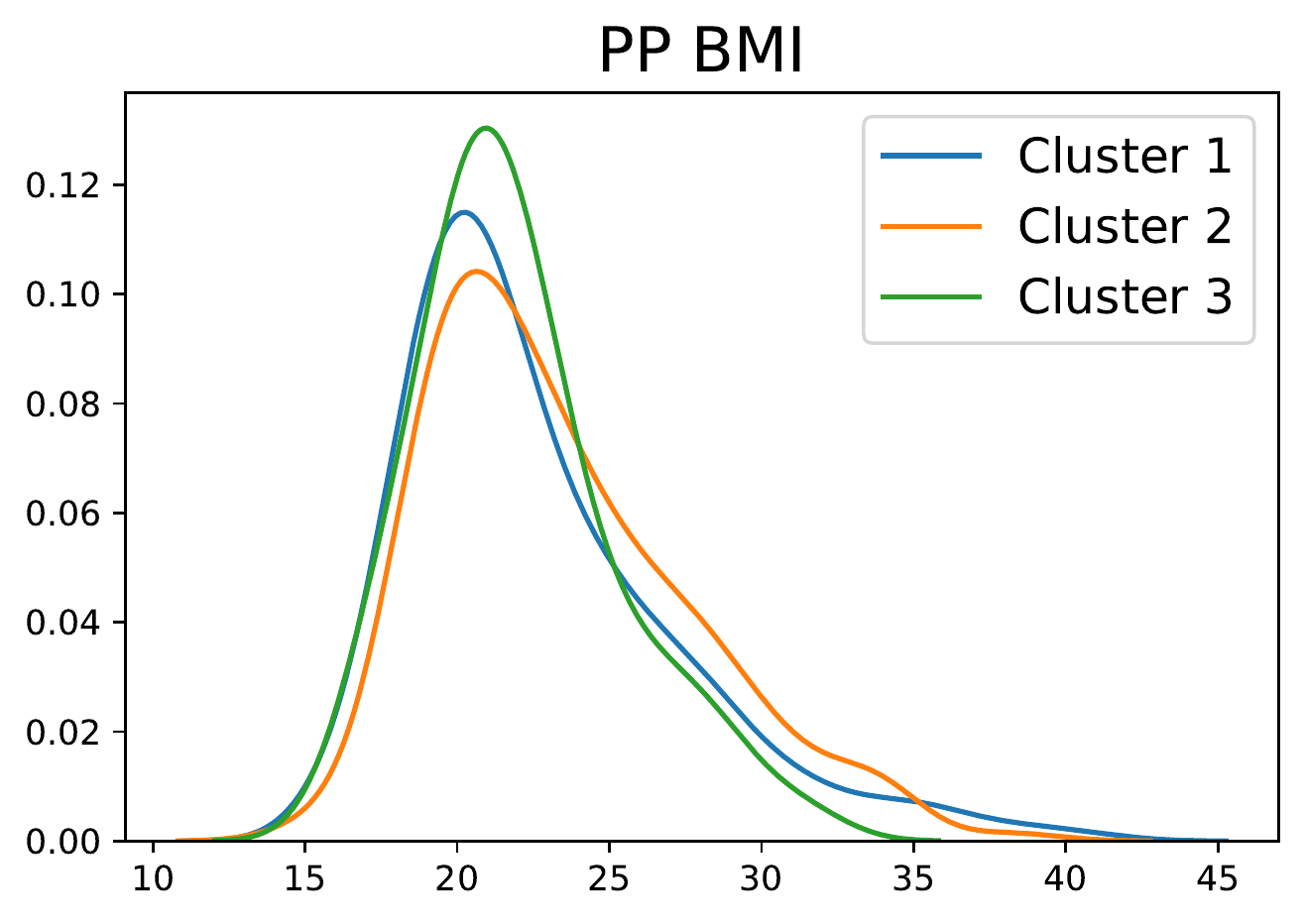}
        \caption{0.07, 0.55, 0.06}
    \end{subfigure}%
    \begin{subfigure}{0.3\linewidth}
        \includegraphics[width=\linewidth]{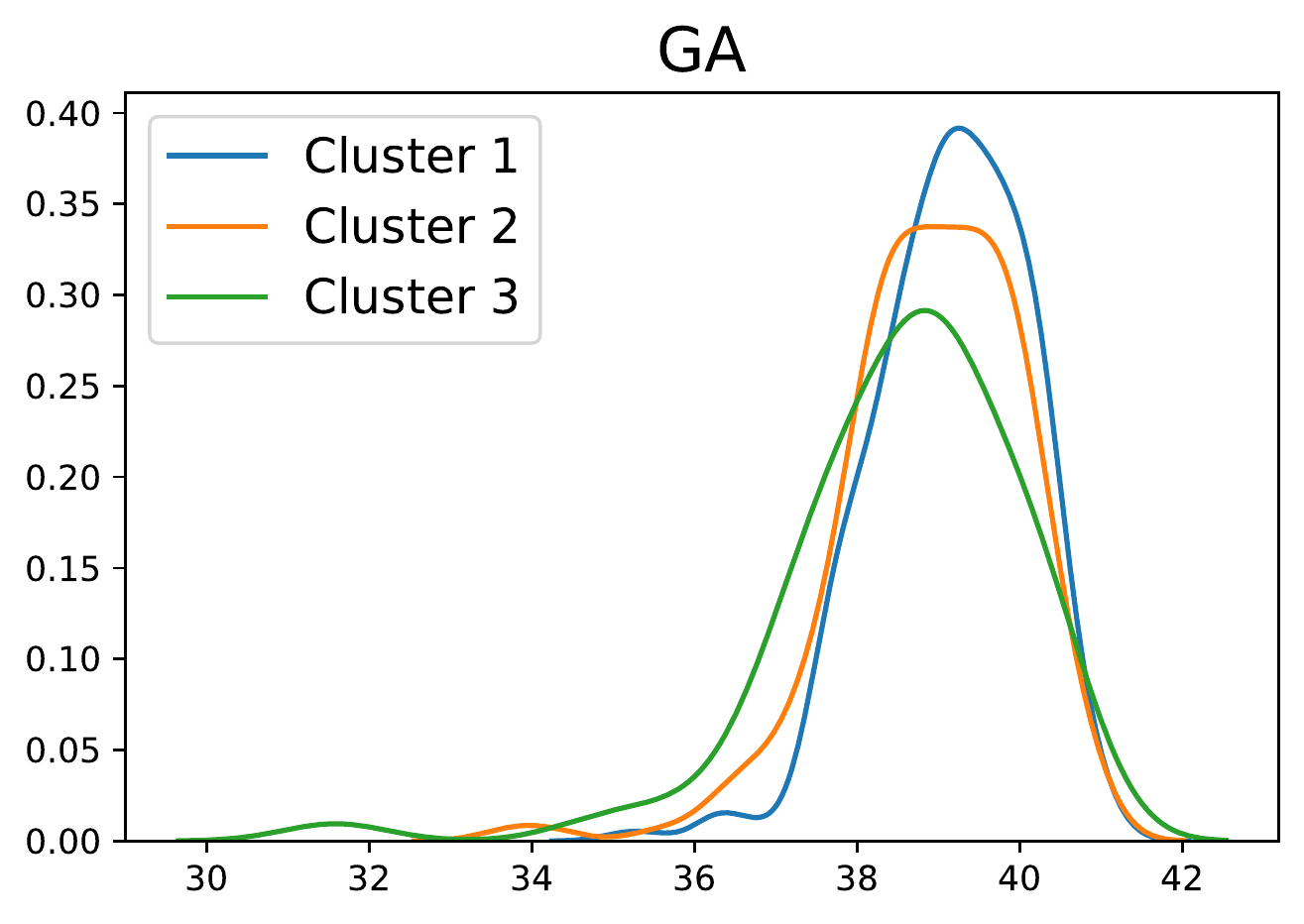}
        \caption{0.07, 0.01, 0.22}
    \end{subfigure}

    \begin{subfigure}{0.3\linewidth}
        \includegraphics[width=\linewidth]{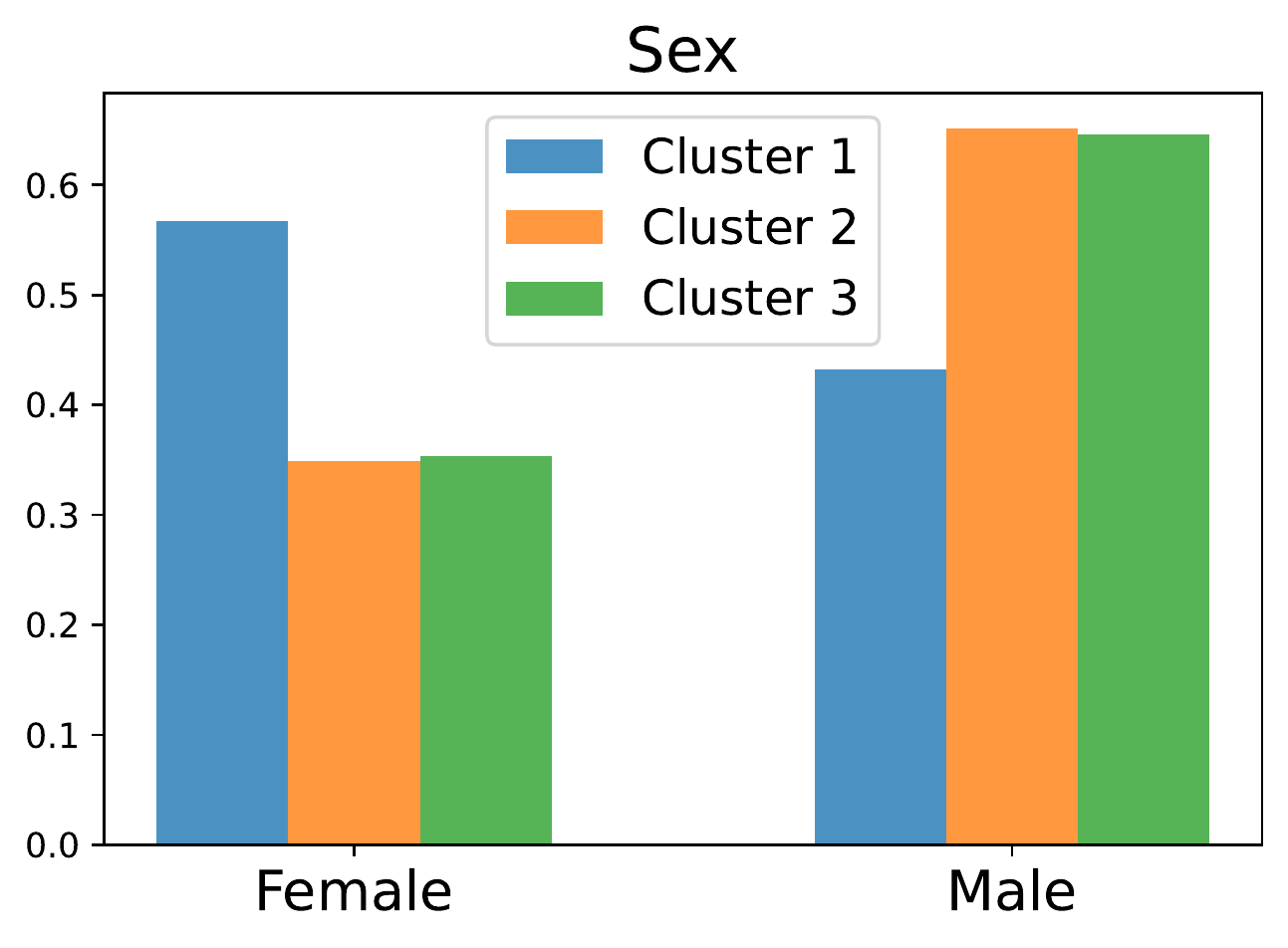}
        \caption{0.00, 0.00, 0.98}
    \end{subfigure}%
    \begin{subfigure}{0.3\linewidth}
        \includegraphics[width=\linewidth]{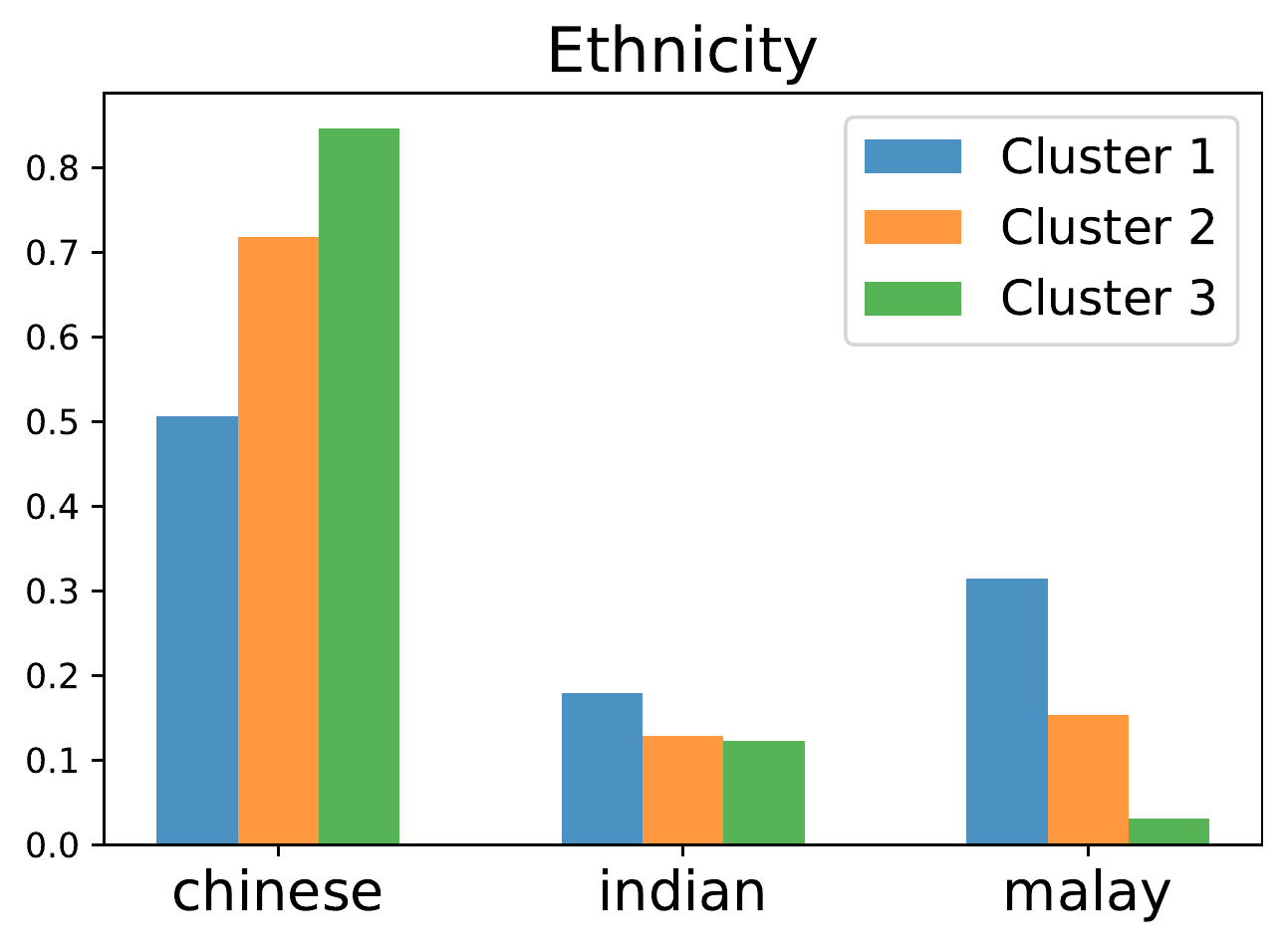}
        \caption{0.00, 0.00, 0.00}
    \end{subfigure}%
    \begin{subfigure}{0.3\linewidth}
        \includegraphics[width=\linewidth]{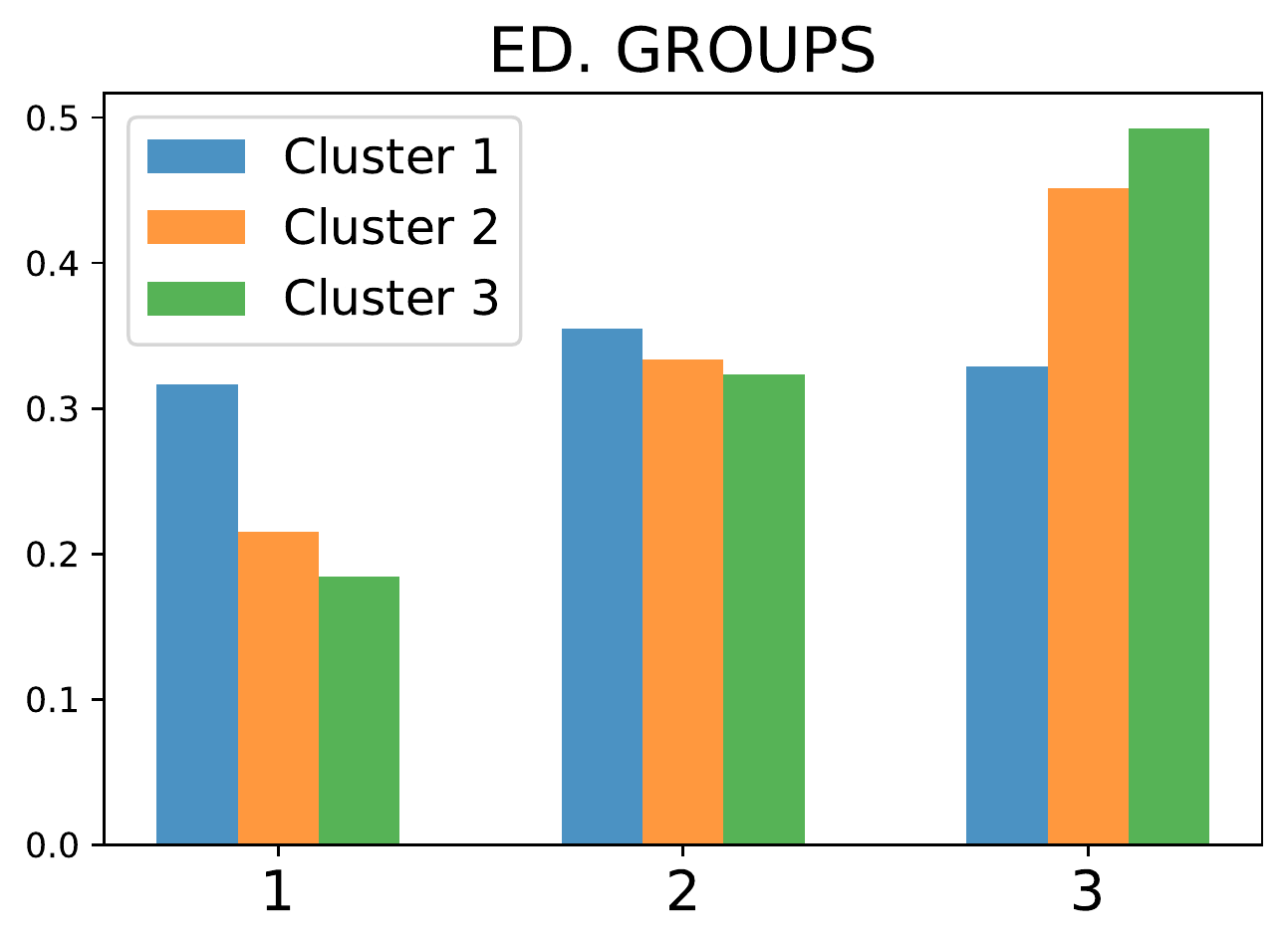}
        \caption{0.00, 0.00, 0.40}
    \end{subfigure}

    \caption{Empirical distribution of the covariates in each cluster. The three numbers below each plot represent the p-values for the homogeneity tests for covariates in clusters (1, 2), (1, 3) and (2, 3), respectively.}
    \label{fig:test_homo_cov}
\end{figure}

As already mentioned, the main three clusters could differ either in the responses or in the covariates (or both). To better understand what discriminates the three main clusters, we perform homogeneity tests for the equality in distribution of both responses and covariates in the different clusters. The results should be considered as a descriptive tool. 
In particular, for the responses we consider the  data on  both height and weight at each visit separately and test the equality of the distributions for each pair of clusters. For each of the covariates, we test the equality of their  distributions in each possible pair of clusters. For the response variables and continuous covariates, we employ the Kolmogorov-Smirnov (KS) test for equality in distribution and the Pearson's chi- squared test of homogeneity for the categorical covariates. 
Table~\ref{tab:p_vals} reports the p-values associated to the KS test for the responses, while Figure~\ref{fig:test_homo_cov} shows the cluster specific empirical distribution of the covariates.
From Table~\ref{tab:p_vals} and Figure~\ref{fig:test_homo_cov}, it is clear that clusters 2 and 3 (second and third rows in Figure~\ref{fig:clusters}, respectively) are similar in terms of both responses at each time point. However, Figure~\ref{fig:test_homo_cov} (bottom row) suggests that the three main clusters cannot be \textit{explained} only in terms of \textit{ethnicity}, even though cluster 3 contains  almost exclusively Chinese children.

\begin{table}[ht]
\centering
\begin{tabular}{l || c | c | c || c | c | c | c}
& \multicolumn{3}{c ||}{Height} & \multicolumn{3}{c}{Weight} \\
Clusters &     (1, 2) & (1, 3) & (2, 3) & (1, 2) & (1, 3) &  (2, 3) \\
\midrule
$t=1$  & \textbf{0.023} & \textbf{0.000} & \textbf{0.025} & \textbf{0.002} & \textbf{0.296} & 0.606 \\
$t=2$  & \textbf{0.000} & \textbf{0.023} & 0.999 & \textbf{0.000} & \textbf{0.000} & 0.785 \\
$t=3$  & \textbf{0.000} & \textbf{0.003} & 0.797 & \textbf{0.000} & \textbf{0.013} & 0.815 \\
$t=4$  & \textbf{0.000} & \textbf{0.000} & \textbf{0.000} & \textbf{0.000} & 0.253 & 0.620 \\
$t=5$  & \textbf{0.046} & \textbf{0.004} & \textbf{0.044} & \textbf{0.000} & 0.197 & 0.386 \\
$t=6$  & \textbf{0.000} & 0.051 & 0.701 & \textbf{0.000} & 0.241 & 0.254 \\
$t=7$  & \textbf{0.003} & 0.113 & 0.878 & \textbf{0.000} & 0.431 & 0.375 \\
$t=8$  & \textbf{0.000} & 0.072 & 0.733 & \textbf{0.000} & 0.210 & 0.718 \\
$t=9$  & \textbf{0.000} & 0.106 & 0.984 & \textbf{0.000} & 0.196 & 0.715 \\
$t=10$  & \textbf{0.000} & 0.112 & 0.869 & \textbf{0.000} & 0.341 & 0.717 \\
$t=11$ & \textbf{0.000} & 0.213 & 0.726 & \textbf{0.000} & 0.244 & 0.854 \\
$t=12$ & \textbf{0.000} & 0.165 & 0.877 & \textbf{0.000} & 0.125 & 0.932 \\
$t=13$ & \textbf{0.000} & 0.179 & 0.993 & \textbf{0.000} & \textbf{0.042} & 0.811
\end{tabular}
\caption{P-values of the homogeneity tests for the equality in distribution at every visit for each pair of clusters, considering height and weight. Bold numbers correspond to p-values lower than 5\%}
\label{tab:p_vals}
\end{table}

\begin{figure}[ht]
\centering
\includegraphics[width=0.9\linewidth]{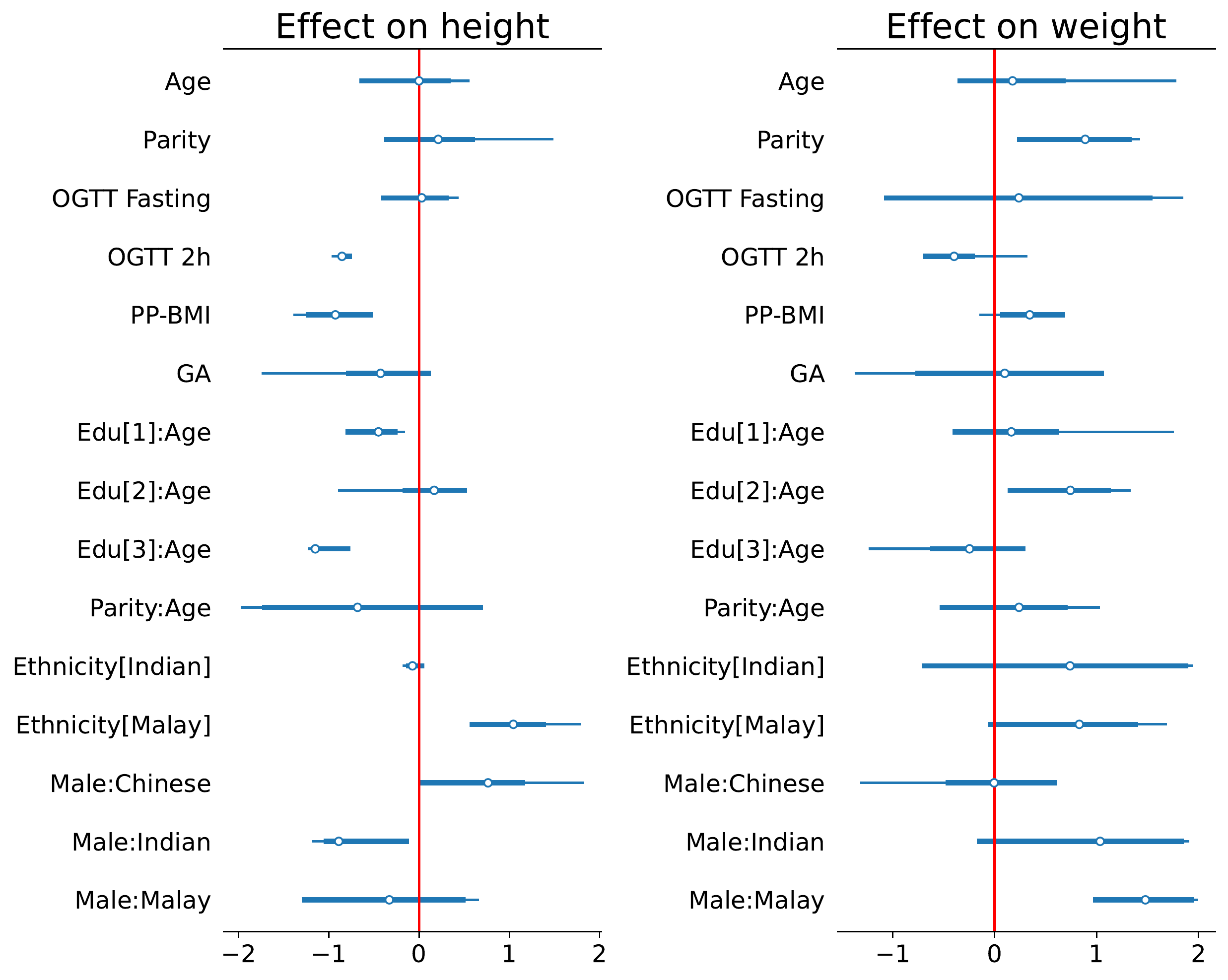}
\caption{
Posterior credible intervals of the regression coefficients in $\Gamma$ for
the height  (left plot) and weight of the children (right plot).
Thin lines correspond to $95\%$ credible intervals, while thick lines to
$80\%$ credible intervals.}
\label{fig:gamma_hdi}
\end{figure}

Next we consider the two parameters in $B$, i.e. the regression parameters for the square root of time $t$ for the two responses; see \eqref{eq:lik}. The posterior means are $5.55, 0.96$, respectively, with  marginal standard deviations $0.02, 0.01$, thus indicating a non-negligible growth trend for both  height and weight, as expected.
Figure~\ref{fig:gamma_hdi} displays  posterior credible intervals for all the parameters in $\Gamma$
defined in \eqref{eq:lik}, that is the regression coefficients corresponding to the  time-homogeneous  covariates.
The reference group for the  categorical covariates has been set such that the baseline level is for a Chinese female child; see Section~\ref{sec:EDA}. 
Covariates such as
\textit{OGTT 2h}, \textit{ppBMI}, the interaction between education and age, ethnicity (Malay) and the interaction between sex and ethnicity have the strongest effects on  height. 
On the other hand, \textit{parity}, \textit{OGTT 2h}, \textit{ppBMI}, the interaction between education and age (but only the second level of education) and the interaction between sex and ethnicity have a  strong association with weight. It is clear from  Figure~\ref{fig:gamma_hdi} that most of the posterior mass for the marginal distribution of 
\textit{ethnicity} is concentrated on positive values. 
Correcting for the autoregressive effect, we see that
\textit{ethnicity} 
might impact obesity as   Indian and Malay children are characterised by a larger posterior expected weight, combined in some cases with a lower posterior expected height.
Moreover, also correcting for the autoregressive effect,
our analysis shows that the posterior expected height of a  Chinese male child  is larger  than the reference (Chinese female child). Similar comments can be made, for instance, regarding  Indian male children being smaller than Indian female children, ans so on.

Mother's age and gestational age do not have a strong effect on the child's height and weight, though this might be due to the fact that  these variables are associated with ethnicity; see Figure~\ref{fig:cat_vs_num}. It is known from the literature that increasing parity is associated with increasing neonatal adiposity in Asian and Western populations  \citep[see][]{tint2016abdominal}; this is confirmed by the marginal posterior distribution of the parameter corresponding to the effect of \textit{parity} on weight in Figure~\ref{fig:gamma_hdi}.

The covarates $z_i$'s play also a key role in the definition of the
stick-breaking prior  as seen from \eqref{eq:logit}. To assess if the proposed  covariate-driven stick-breaking prior provides significant advantages  over more standard models, 
we compare it with three possible competitors.
The first one is the parametric  version of our model obtained by setting $H=1$.   The second model assumes a truncated Dirichlet process as a prior for $\Phi_i$'s, with $H=50$, similarly to what is done in Section~\ref{sec:simulation}.
Moreover, as  third competitor prior,  we assume that the $\Phi_i$'s take into account information from the time-homogeneous covariates through the atoms $\Phi_{0h}$'s.
Specifically, the prior for $\Phi$ is specified as in \eqref{eq:stick}, but for each $h=1, \ldots, H$ we define a matrix $\Omega_h \in \mathbb{R}^{k^2 \times q}$ and we let $vec(\Phi_{0h}(\bm z_i)) =: \varphi_{0h}(\bm z_i) = \Omega_h \bm z_i$. 
The weights $\bm w$ in \eqref{eq:stick} do not depend on the value of $\bm z_i$ (i.e., $w_h(\bm z_i) = w_h$) and follow a truncated Dirichlet process prior with $H=50$.
This model can be seen as a finite dimensional approximation of the Linear-DPP in \cite{deiorio2004anova}.

For all the models, we match the prior for $B$, $\Gamma$, $\Sigma$ and, when possible also $H$ and the marginal prior distribution of $\Phi_{0h}$. For the Linear-DPP we assume that the vectorization of the $\Omega_h$'s are independent and identically distributed multivariate Gaussian random variables  with mean zero and identity covariance matrix. Since the full conditional distribution of the $\Omega_h$'s in the case of the Linear-DPP prior does not belong to a known parametric family, we  update them  via an adaptive Metropolis Hastings \citep{andrieu2008tutorial} step.

The different models are compared using widely applicable information
criterion    \citep[WAIC,][]{watanabe2013widely}. Higher  values of WAIC correspond to better predictive performances. We marginalize the missing values from the predictive distribution  of the response trajectory and consider just the marginal predictive distribution for the non-missing values.
We found that WAIC is equal to $-3.4 \times 10^6$  for the Linear-DDP, $-6.7 \times 10^5$ 
for the parametric model, $-3.9 \times 10^5$ for the DP model and $-3.4 \times 10^5$ for our model, confirming that our model performs   better than the competitors. 
 Moreover, we report that  the MCMC algorithm for the Linear-DDP requires a much larger number of burn-in iterations ($10^5$ vs. $10^4$) than the other models to reach satisfactory convergence, and that the expected 
 number of cluster a posteriori in the Linear-DPP is around 42.
It is then clear that (i) assuming linear dependence of the fixed-time covariates in the autoregressive parameters matrices $\Phi_i$ does not give good predictive fit (or at least not better than our model), and that (ii) adding covariate information in the stick-breaking prior improves the prediction performance.

\section{Summary}\label{sec:discussion}

 The aim of this manuscript is to cluster children according to obesity growth patterns.
Obesity is an epidemic, increasingly affecting children.
Overweight or obesity in childhood may be critical as they often persist into adulthood due to both physiological and behavioral factors.

Motivation for our study stems from a child growth dataset. To analyze these data we developed a Bayesian
nonparametric VAR joint model for height and weight profiles for
these children.
One key aspect behind the modeling choice was to cluster the corresponding joint
time-evolving profiles using the available covariate information. The model
features a logit stick-breaking construction that can accommodate covariate
dependence in the mixture weights. This allows us to relate certain baseline
conditions of these children, such as sex or ethnicity, to obesity patterns. Ethnic differences in obesity are of interest as they could be due to genetic factors, dietary intake, cultural or socioeconomic factors. The analysis allowed us to identify important clusters of children that are
characterized by differences in the trajectories or in the covariates or both.

Posterior inference was carried out by means of an efficient posterior simulation
that exploits recently developed results on logit stick-breaking priors, which
facilitates postulating covariate dependence in the mixture weights. For this implementation we chose to fix a sufficiently large number of components from
which we focused on the number of these that were actually occupied (we referred
to these as \textit{clusters}). The results obtained were compared
against competitor models, and we found that our approach provides superior performance as measured by standard quantities such as the WAIC.

\FloatBarrier

\section*{Acknowledgments}
This work was partially funded by grant FONDECYT 1180034.

\appendix

\section{Further plots}
\label{sec:fur_plots}
\begin{figure}[!ht]
\centering
\includegraphics[width=\textwidth,height=0.7\textheight,keepaspectratio]{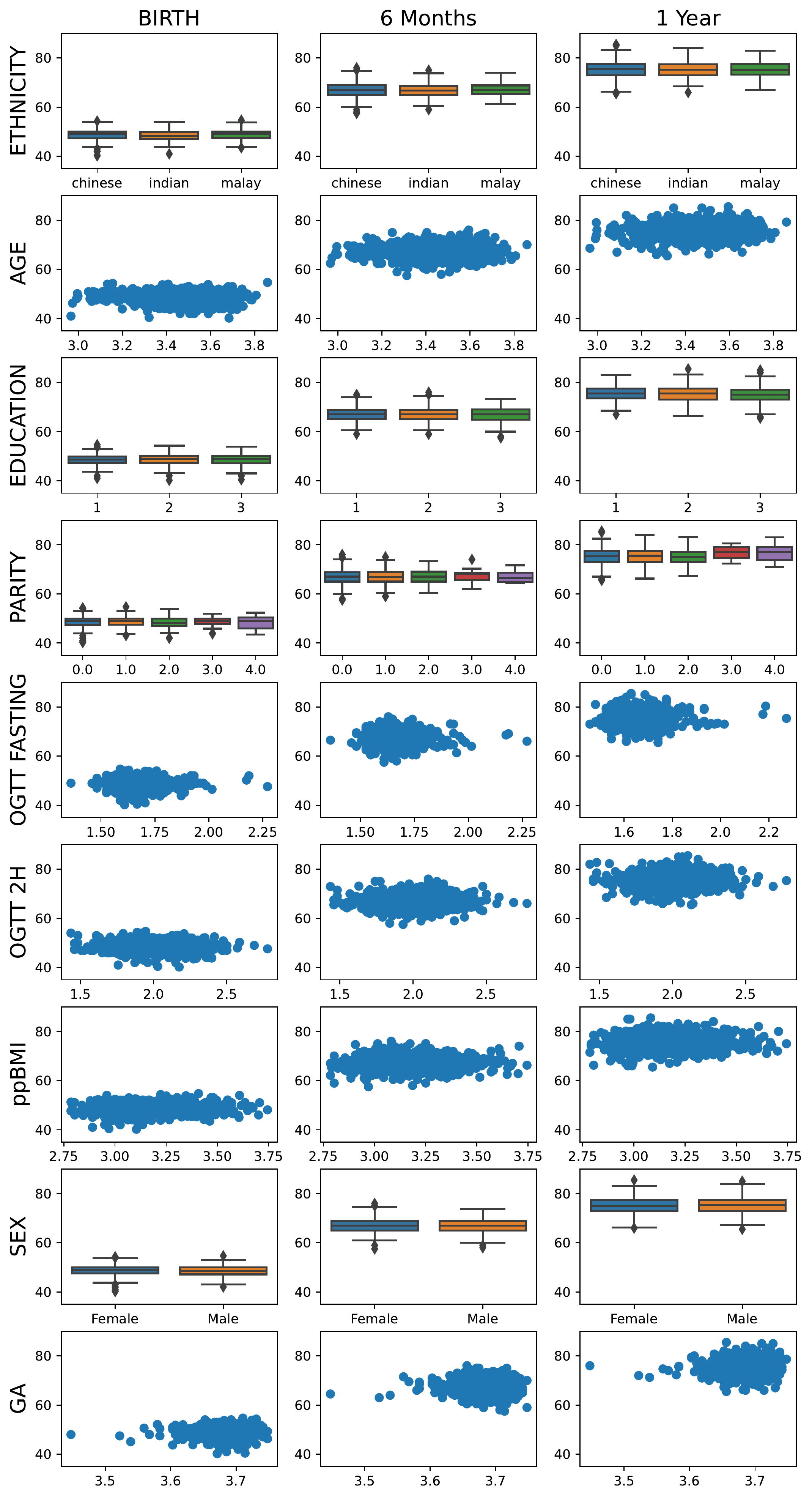}
\caption{Scatterplots of covariates against the height at birth (left column), at six months of age (center) and at one year of age (right column).}
\label{fig:scatter_height}
\end{figure}

\begin{figure}[!ht]
\centering
\includegraphics[width=\textwidth,height=0.7\textheight,keepaspectratio]{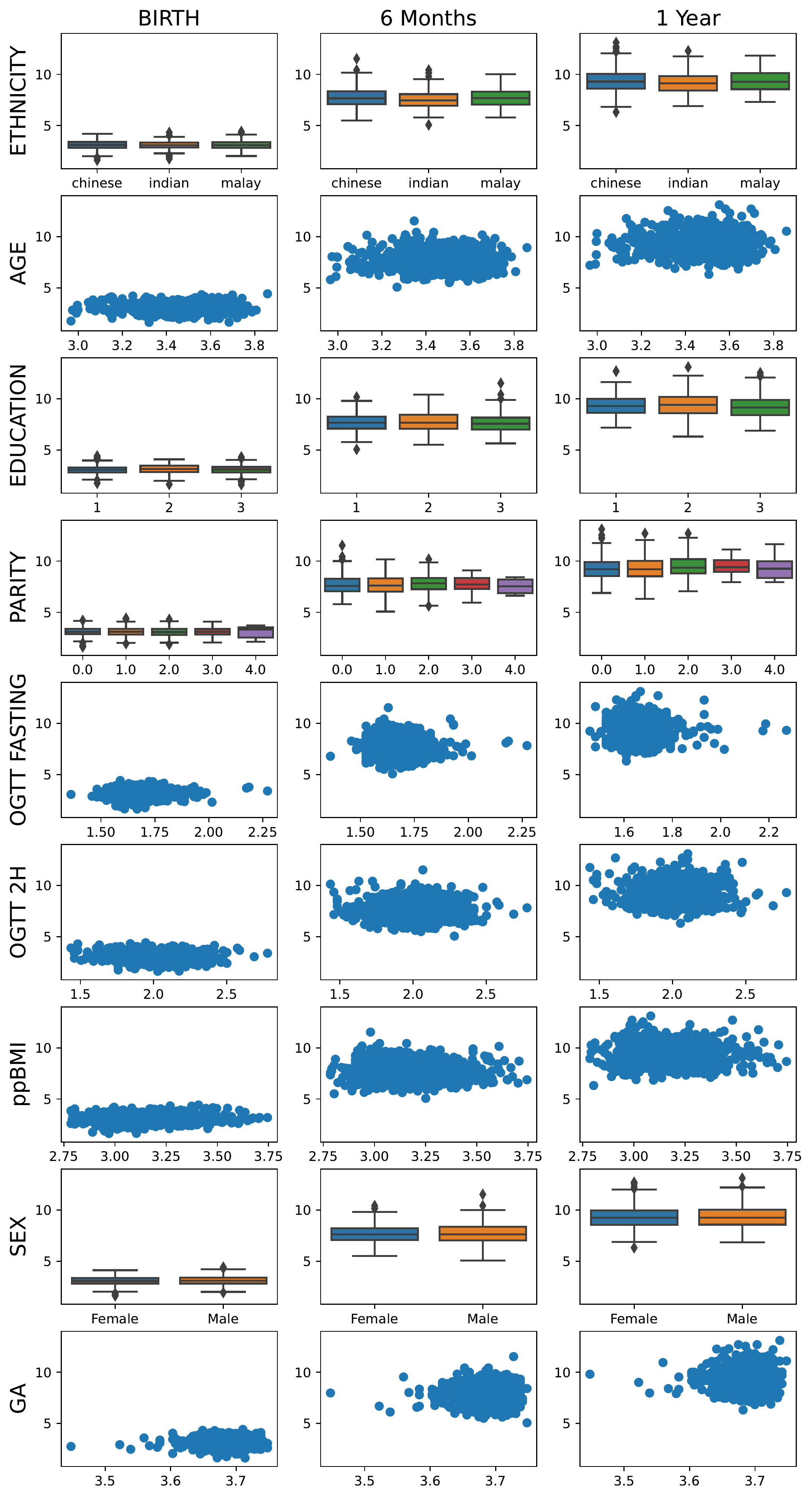}
\caption{Scatterplots of covariates against the weight at birth (left column), at six months of age (center) and at one year of age (right column).}
\label{fig:scatter_weight}
\end{figure}

We show the scatterplots   of the responses (height and weight) at time $t=0,1,2$ versus all continuous covariates at the baseline of the dataset on obesity for Children in Singapore. When the covariate we consider is discrete, scatterplots are replaced by boxplots. The left column of Figure~\ref{fig:scatter_height} reports scatterplots or boxplots of the height at time $t=0$, while the left column of Figure~\ref{fig:scatter_weight} reports similar plots for the weight at time $t=0$. The central and right columns of 
Figures~\ref{fig:scatter_height} and \ref{fig:scatter_weight} display the same plots of the responses at time $t=1$ and 2.  

\section{The Gibbs sampler}
\label{sec:gibbs}
Posterior inference for our logit stick-breaking model \eqref{eq:lik}-  \eqref{eq:atoms_prior_II} is carried out using a Gibbs sampler algorithm, with full conditionals outlined below. The joint distribution of data and parameters is described here
\begin{flalign}
   \Law(\bm Y_1, \dots, \bm Y_N, B, \Gamma, \Sigma, {\Phi}_1,\ldots,{\Phi}_N) &= \prod_{i=1}^{N} \Law(Y_{i1}, \dots, Y_{iT_i}|\bm b,\bm\gamma,\Sigma, \Phi_1, \ldots, \Phi_n) \\\nonumber
   & \times \pi(\bm b) \times \pi(\bm \gamma) \times \pi(\Sigma) \times \pi(\Phi_1, \dots, \Phi_N | \bm z_1, \dots, \bm z_N)
\end{flalign}

In what follows, \virgolette{rest} refers to to the data and all parameters except for the one to the left of \virgolette{$\mid$}.
Moreover we adopt the matrix notation or the vector one for all parameters interchangeably.

As in \cite{ishwaran_2001_gibbs}, to sample from the stick-breaking prior on $\Phi_i$, as it is standard, we use cluster indicator latent variables, that will be indicated by $G_i$.

\begin{enumerate}

\item The full-conditional for the parameters $\bm b = \text{vec}(B)$ can be obtained by noticing that using the following change of variable
\[
 \bm{y_{it}} - \Phi_i \bm{y_{it-1}} - \Gamma \bm{z_i} = B \bm{x_{it}} + \bm{\epsilon_{it}}
\]
we recover the standard expression of Bayesian multivariate linear regression, let $\bm{w_{it}} = \bm{y_{it}} - \Phi_i \bm{y_{it-1}} - \Gamma \bm{z_i}$. We have:
\begin{equation*}
	\bm{w_{it}} = \bm{x_{it}}^T B^T + \bm{\epsilon_{it}}.
\end{equation*}
Using standard techniques, calling
\begin{equation*}
	\bm W = 
	\begin{bmatrix}
	\bm w_{1,1} \\ \vdots \\ \bm w_{i, T_1} \\ \vdots \\ \bm w_{N, 1} \\ \vdots \\ \bm w_{N,T_N}  
	\end{bmatrix} \quad 
	X = \begin{bmatrix}
	\bm x_{1,1} \\
	\vdots \\
	\bm x_{1, T_1} \\
	\vdots \\
	\bm x_{N, 1}\\ 
	\vdots \\
	\bm x_{N, T_N}
	\end{bmatrix}
\end{equation*}
We can write the system in vector form as:
\begin{equation*}
 \bm W = X B^T + \bm E,
\end{equation*}
where $W, E$ are $[\sum_{i=1}^N T_i \times k ]$ matrices and $X$ is $[\sum_{i=1}^N T_i \times p]$.
By standard multivariate regression theory we have that
\begin{flalign*}
	\bm b | X, W, \Sigma &\sim \calN \left( \widetilde{\mu_b}, \widetilde{\Sigma_b} \right) \\\nonumber
	\mu_b &=(\Sigma^{-1} \otimes X^T X + \Sigma_b^{-1})^{-1} \left( (\Sigma^{-1} \otimes X^T X) \hat{\beta} + \Sigma_b^{-1} \widetilde{\beta_0} \right) \\\nonumber
	\widetilde{\Sigma_b} &= \Sigma^{-1} \otimes X^T X + \Sigma_b^{-1},
\end{flalign*}
where $\hat{\beta}$ is the standard frequentist estimate:
\begin{equation*}
	\hat{\beta} = (X^T X)^{-1} X^T \bm W.
\end{equation*}
We thus obtain:
\begin{equation}
	\Law(\bm b | \text{rest}) = \calN(\widetilde{\mu_b}, \widetilde{\Sigma_b})
\end{equation}

\item Analogously to what we did in the previous step, the law of $\bm \gamma$ can be deducted from standard Bayesian multivariate regression theory after a suitable change of variable:

\begin{equation*}
	\bm y_{it} - \Phi_i \bm y_{it-1} - B \bm x_{it} = \Gamma \bm z_i + \bm \epsilon_{it}
\end{equation*}

We thus recover the same equations as in the previous section.

\item To sample from $\Law (\Sigma | \text{rest} )$ we analyze the full conditional (to simplify the notation we impose $\bm y_{i0} = \bm 0$ for all $i$'s):
\begin{flalign*}
\Law(\Sigma^{-1} | - ) &\propto \prod_{i=1}^N \prod_{t=1}^{T_i} \frac{1}{\Det{2 \pi \Sigma}^{\frac{1}{2}}} \exp \left( -\frac{1}{2} (\bm y_{it} - \Phi_i \bm y_{it-1} - B \bm x_{it} -\Gamma \bm z_i)^T \Sigma^{-1} (\bm y_{it} - \Phi_i \bm y_{it-1} - B \bm x_{it} -\Gamma \bm z_i) \right)  \\\nonumber
& \times \frac{\Det{\Sigma_0}^{\frac{\tau}{2}}}{2^{\frac{\tau k}{2}} \Gamma_k \left(\frac{\tau}{2} \right)} \Det{\Sigma}^{-\frac{\tau + k + 1}{2}} \exp \left(-\frac{1}{2} \tr(\Sigma_0 \Sigma^{-1}) \right) \\\nonumber
&\propto \frac{\Det{\Sigma}^{-\frac{\tau + k + 1}{2}}}{\Det{2\pi\Sigma}^{+\frac{1}{2} \sum_{i=1}^{N} T_i}} \exp \left(-\frac{1}{2} E \right).
\end{flalign*}
By using the trace trick, circularity of the trace and linearity of the trace operator we get that
\begin{equation*}
	E = \tr \left(\left(\sum_{i=1}^{N} \sum_{t=1}^{T_i} (\bm y_{it} - \Phi_i \bm y_{it-1} - B \bm x_{it} -\Gamma \bm z_i) (\bm y_{it} - \Phi_i \bm y_{it-1} - B \bm x_{it} -\Gamma \bm z_i)^T + \Sigma_0 \right)\Sigma^{-1} \right).
\end{equation*}

We can deduce that $\Law(\Sigma \mid \text{rest}) = IW(\tilde{\nu}, \widetilde{\Sigma_0})$  with parameters
\begin{flalign}
\tilde{\nu} & = \nu + \sum_{i=1}^{N} T_i \\\nonumber
\widetilde{\Sigma_0} &= \sum_{i=1}^{N} \sum_{t=1}^{T_i} (\bm y_{it} - \Phi_i \bm y_{it-1} - B \bm x_{it} -\Gamma \bm z_i) (\bm y_{it} - \Phi_i \bm y_{it-1} - B \bm x_{it} -\Gamma \bm z_i)^T + \Sigma_0.
\end{flalign}

\item The component indicator variables are sampled considering the usual change of variables 
\[
	\bm w_{it} = \Phi_i(\bm z_i) \bm y_{it-1} + \bm \epsilon_{it},
\]
where $\bm w_{it} = \bm y_{it} - B \bm x_{it} - \Gamma z_i$.
We have that:
\begin{flalign}
\label{eq:weights_Gi}
	P(G_i=h | \text{rest}) & \propto P(G_i=h) f(\bm w_{i1}, \dots, \bm w_{iT_i} | G_i=h) \\\nonumber
	& \propto P(G_i=h  \times f(\bm w_{i1} | G_i=h, \text{rest}) \prod_{t=2}^{T_i} f(\bm w_{it} | \bm y_{it-1}, \text{rest}) \\\nonumber
& \propto \nu_h (\bm z_i) \prod_{l=1}^{h-1} \left( 1 - \nu_l (\bm z_i) \right) \times \calN(\bm w_{i1}; \bm B x_{i1} + \Gamma z_i, \Sigma) \prod_{t=2}^{T_i} \calN(\bm w_{it}; \Phi_{0h} \bm y_{it-1}\bm B x_{it} + \Gamma z_i, \Sigma).
\end{flalign}

Thus the conditional distribution of $G_i$ is a discrete distribution with weights as in \eqref{eq:weights_Gi}.

\item For each cluster-specific $\Phi_{0h}$ we have that, for the $i$'s such that $G_i = h$:
\[
	\bm y_{it} = \Phi_{0h} \bm y_{it-1} + B x_{it} + \Gamma z_i + \epsilon_{it}.
\]

Defining:
\begin{equation*}
	\bm Y = 
	\begin{bmatrix}
	\bm y_{11} - B \bm x_{11} - \Gamma \bm z_{1} \\
	\vdots \\ 
	\bm y_{1 T_1} - B \bm x_{1 T_1} - \Gamma \bm z_{1} \\
	\vdots \\
	\bm y_{N_i1} - B \bm x_{N_i1} - \Gamma \bm z_{N_i} \\
	\vdots \\
	\bm y_{N_i T_{N_i}} - B \bm x_{N_i T_{N_i}} - \Gamma \bm z_{N_i}
	\end{bmatrix}
	 \quad 
	X = \begin{bmatrix}
	\bm y_{10} \\
	\vdots \\
	\bm y_{1 T_1 -1 } \\
	\vdots \\
	\bm y_{N_i 1} \\
	\vdots \\
	\bm y_{N_i T_i -1}
	\end{bmatrix}
\end{equation*}
where the $\bm y_i$s have been selected such that they belong to cluster $h$, we have the following Seemingly Unrelated Representation:
\begin{equation}
	\bm Y = X \Phi_{0h}^T + E.
\end{equation}
Thus, we can recover the full conditional for $\varphi_{0h} := \vvec \left(\Phi_{0h}\right) $ using standard Bayesian multivariate regression theory. In particular we have that:
\begin{flalign}
	\varphi_{0h} | Y, X, \Sigma & \sim \calN (\mu_{0h}, \Sigma_{0h}) \\\nonumber
	\Sigma_{0h} &= \Sigma^{-1} \otimes X^T X + V_0^{-1} \\\nonumber
	\mu_{0h} &= \Sigma_{0h}^{-1} \left((\Sigma^{-1} \otimes X^T X) \widehat{\varphi_{0h}} + V_0^{-1} \varphi_{00} \right),
\end{flalign}

where $\widehat{\varphi_{0h}} = (X^T X)^{-1} X^T Y$ is the frequentist estimation.

\item Since the update of $\alpha_h$ is independent of the AR model, we can simply refer to \cite{rigon_durante_logit} where a latent variable $\omega_{ih}$ is introduced. Defining $\rho_{ih} | \bm z_i \sim \calB(\nu_h(\bm z_i))$, the couple $(\omega_{ih}, \rho_{ih})$ is updated as in \cite{polson_etal_jasa_2013} from a P\'olya-Gamma distribution.

\item As the joint law does not depend from the parameters $\Phi_{00}, V_{0}$ except for the prior specification of $\Phi_{0h}$, we can update them using a Normal-Normal-inverse-Wishart scheme as follows:
\begin{flalign}
 \varphi_{0h} | \varphi_{00}, V_0 & \iid \calN(\varphi_{00}, V_{0}) \\\nonumber
 \varphi_{00} | \varphi_{000}, V_0, \lambda_0 & \sim \calN \left(\varphi_{000}, \frac{1}{\lambda_0} V_0 \right) \\\nonumber
 V_0 | V_{00}, \tau_{0} & \sim \invWish (V_{00}, \nu_0), \\\nonumber
\end{flalign}
From this we have that:
\begin{flalign}
 \varphi_{00} | V_0, \varphi_{01}, \dots \varphi_{0H} & \sim \calN \left( \frac{H \overline{\varphi_0} + \lambda \varphi_{000}}{H + \lambda}, \frac{1}{H + \lambda} V_0 \right) \\\nonumber
 V_0 | \varphi_{01}, \dots \varphi_{0H} & \sim \invWish \left( V_{00} + H S + \frac{H \lambda}{H + \lambda}(\overline{\varphi_0} - \varphi_{000}) (\overline{\varphi_0} - \varphi_{000})^T, H + \nu_0 \right) \\\nonumber
 \overline{\varphi_0} &= \frac{1}{H} \sum_{h=1}^H \varphi_{0h} \\\nonumber
 S &= \frac{1}{H} \sum_{h=1}^H (\varphi_{0h} - \overline{\varphi_0}) (\Phi_{0h} - \overline{\varphi_0})^T
\end{flalign}

\end{enumerate}

An iteration of our Gibbs samples consists in sampling from the full conditionals described in steps 1. through 7. above, iteratively. Moreover, if there are missing responses as in the case of the application, at each iteration, before step 1., we sample the missing responses from their full conditional as described  below.

\subsection{Sampling missing responses}
\label{sec:missingresponses}

We start by deriving the joint law of the vector $\bm y_i = (\bm y_{i1}, \ldots, \bm y_{iT_i})$, given $\Phi_i, B, \Gamma$ and $\Sigma$.
Consider the simplified VAR model, for a single patient (we drop the index $i$).
\begin{equation}
    \label{eq:simplified_var}
    \bm y_1 = \epsilon_1, \quad \bm y_t | X_{t-1} = \Phi \bm y_{t-1} + \epsilon_t.
\end{equation}
By expressing the joint law as $\Law(\bm y_1, \ldots, \bm y_T) = \Law(\bm y_1) \Law(\bm y_2 | \bm y_1)\dots \Law(\bm y_T | \bm y_{T-1})$ and through some basic linear algebra, we 
can derive that the vectorization of $(\bm y_1, \ldots, \bm y_T)$ is a jointly normal
random vector with zero mean.
The precision matrix $\widetilde{\Sigma}^{-1}$ of the normal distribution has a blocked structure made of
$T \times T$ blocks, each of which is an $r \times r$ matrix.
The $(i, j)$-th block equals to:
\begin{equation}\label{eq:missing_val_prec}
    \widetilde{\Sigma}^{-1}_{i, j} =
    \begin{cases}
        (I + \Phi)^T \Sigma^{-1} (I + \Phi), & \text{if } i=j< T\\
        \Sigma^{-1}, & \text{if } i=j=T \\
        \Phi^T \Sigma^{-1}, & \text{if } |i-j| = 1 \\
        0 & \text{if } |i-j| > 1
    \end{cases}
\end{equation}

Going back to the full model, it is easy to see that with a change of variable $\bm y_{it} \mapsto \bm y_{i, t} - B x_{i, t} - \Gamma z_i$ we recover the same VAR system in~\eqref{eq:simplified_var}.
Hence, the vectorization of $\bm y_i$ follows a multivariate normal with precision matrix given by~\eqref{eq:missing_val_prec} and mean $\bm \mu$ given by the vectorization 
of $(B\bm x_{i1} + \Gamma z_i, \ldots, B\bm x_{iT_i} + \Gamma z_i)$.

To simulate missing values in $\bm y_i$, we exploit the joint law derived above and the fact that the conditional distributions of entries in a Gaussian random vector are available in close form.
In particular, if there are $k$ missing values in $\bm y_i$, we first apply a permutation matrix $P$ to the vectorization of $\bm y_i$ so that the missing entries are the first $k$ (this will in turn change the mean $\bm \mu$ to $P \bm \mu$ and the covariance matrix to $P^T \widetilde{\Sigma} P$).  Then, using notation $\bm x^{:k}$  and $\bm x^{k:}$ for the first $k$ elements of vector $\bm x$ and the elements $k+1, \ldots$ respectively, and notation $A^{:k, \ell:}$ for a matrix $A$ analogously, where the first index denotes the rows and the second index denotes the columns, we have that:
\[
    (P\bm y_i)^{:k} \mid  (P\bm y_i)^{k:} \sim \mathcal{N}_{k}(\overline{\bm \mu}, \overline{\Sigma}),
\]
where 
\[
    \overline{\bm \mu} = [P(B \bm x_i + \Gamma \bm z_i)]^{:k} + [P^T \widetilde{\Sigma} P]^{k:, :k} \left([P^T \widetilde{\Sigma} P)]^{:k, :k} \right)^{-1} (P\bm y_i^{k:} - P(B \bm x_i + \Gamma \bm z_i)]^{k:})
\]
and 
\[
    \overline{\Sigma} = [P^T \widetilde{\Sigma} P]^{k:, :k} \left([P^T \widetilde{\Sigma} P)]^{:k, :k} \right)^{-1} [P^T \widetilde{\Sigma} P]^{:k, k:}
\]
See Proposition 3.13 in \cite{eaton1983multivariate} for a proof.  

\bibliography{references_paper}
\bibliographystyle{ba}

\end{document}